\documentclass[useAMS,usenatbib]{mn2e}

\usepackage{longtable}
\usepackage{subfigure}
\usepackage{graphicx}
\usepackage{gensymb}
\usepackage{lscape}
\usepackage{afterpage}
\title[Modelling CIRs using realistic spots]{Investigating the origin of cyclical wind variability in hot, 
massive stars - II. Hydrodynamical simulations of co-rotating interaction regions using realistic spot parameters for the O giant $\xi$ Persei}
\author[A. David-Uraz et al.]{A. David-Uraz,$^{1,2,3}$\footnotemark[2] S. P. Owocki,$^{4}$ G. A. Wade,$^{1}$ J. O. Sundqvist,$^{5,6}$ and N. D. Kee$^{4,7}$\\
$^{1}$Department of Physics, Royal Military College of Canada, PO Box 17000, Stn Forces, Kingston, Canada, K7K 4B4\\
$^{2}$Department of Physics, Engineering Physics and Astronomy, Queen's University, 99 University Avenue, Kingston, Canada, K7L 3N6\\
$^{3}$Department of Physics \& Space Sciences, Florida Institute of Technology, Melbourne, FL 32901, USA\\
$^{4}$Bartol Research Institute, University of Delaware, Newark, DE 19716, USA\\
$^{5}$Centro de Astrobiolog\'{i}a, CSIC-INTA, Ctra. Torrej\'{o}n a Ajalvir km. 4, E-28850 Madrid, Spain\\
$^{6}$Instituut voor Sterrenkunde, KU Leuven, Celestijnenlaan 200D, B-3001 Leuven, Belgium\\
$^{7}$Institut f\"{u}r Astronomie und Astrophysik, Universit\"{a}t T\"{u}bingen, Auf der Morgenstelle 10, D-72076 T\"{u}bingen, Germany}
\begin{document}

\date{Accepted 2017 June 12. Received 2017 June 08; in original form 2016 September 23}

\pagerange{\pageref{firstpage}--\pageref{lastpage}} \pubyear{2016}

\maketitle

\label{firstpage}

\begin{abstract}

OB stars exhibit various types of spectral variability historically associated with wind structures, including 
the apparently ubiquitous discrete absorption components (DACs). These features have been proposed to be caused either by magnetic fields or non-radial pulsations.
In this second paper of this series, we revisit the canonical phenomenological hydrodynamical modelling 
used to explain the formation of DACs by taking into account
modern observations and more realistic theoretical predictions.
Using constraints on putative 
bright spots located on the surface of the O giant $\xi$ Persei 
derived from high precision space-based broadband optical photometry obtained with
the Microvariability and Oscillations of STars (\textit{MOST}) space telescope, 
we generate two-dimensional hydrodynamical simulations of
co-rotating interaction regions in its wind. 
We then compute synthetic ultraviolet (UV) resonance line profiles using Sobolev Exact Integration and compare them with
historical timeseries obtained by the International Ultraviolet Explorer (\textit{IUE}) 
to evaluate if the observed behaviour of $\xi$ Persei's DACs is reproduced. Testing three
different models of spot size and strength, we find that the classical pattern of variability can be successfully reproduced for two of them:
the model with the smallest spots yields absorption features
that are incompatible with observations. Furthermore, we test the effect
of the radial dependence of ionization levels on line driving, but cannot 
conclusively assess the importance of this factor. In conclusion, this study 
self-consistently links optical photometry and UV spectroscopy, paving the way to a better understanding of cyclical wind variability
in massive stars in the context of the bright spot paradigm.

\end{abstract}


\footnotetext[2]{E-mail: adaviduraz@fit.edu}

\section{Introduction}

OB stars are known to host various types of wind variability. Most notably, ``discrete absorption components" (DACs), thought
to be ubiquitous \citep{1989ApJS...69..527H} and
observed to migrate through the velocity space of UV resonance lines,
are believed to stem from the presence of large-scale azimuthal density structures in the wind (called ``co-rotating interaction regions", or CIRs; 
\citealt{1986A&A...165..157M}). The most important observational constraint in understanding DACs comes from timeseries of UV 
resonance lines obtained by the International Ultraviolet
Explorer (\textit{IUE}; e.g., \citealt{1986ApJS...61..357P, 1996A&AS..116..257K, 1997A&A...327..281K}).
A key result of those observations is that
the DAC recurrence timescales seem to correlate with the projected rotational velocity ($v \sin i$), suggesting that they are rotationally modulated 
\citep{1988MNRAS.231P..21P}.
%Furthermore, there are a number of DAC properties that can help constrain the conditions in the wind: their maximum depth, the range of velocities that they span,
%as well as their overall morphology. A successful model should allow us to reproduce all of these properties.

Although the physical origin of these structures is unknown, the two main hypotheses to explain them involve magnetic fields
and non-radial pulsations (NRPs). However, both scenarios encounter a number of difficulties in explaining the observed behaviours consistently. 
Indeed, less than 10\% of massive stars harbour detectable magnetic
fields \citep{2014IAUS..302..265W} and those that do usually exhibit large-scale dipolar fields. 
\citet{2014MNRAS.444..429D} have shown that out of a sample of 13 well-studied stars with ultraviolet spectroscopic timeseries showing DACs, none
hosted a detectable dipolar magnetic field. Futhermore, the inferred field strength upper limits excluded any significant influence of undetected
dipolar magnetic fields on the stellar winds. Hence dipolar magnetic fields cannot be responsible for the general phenomenon of DACs. 
On the other hand, the timescales related to DACs are hard to reconcile with typical NRP periods \citep{1999A&A...345..172D}.

Some understanding of the formation of CIRs and DACs can be gained from the phenomenological hydrodynamical modeling carried out
by \citealt{1996ApJ...462..469C} (henceforth
referred to as the ``CO96 model"). Making no physical assumptions about their origin or formation,
it uses \textit{ad hoc} bright spots on the photosphere to drive a locally enhanced outflow, which then leads to rotationally modulated
wind structures. 

Various observations concur that DACs first form at low velocities, before migrating to higher velocities, and therefore that the related structures must exist 
at, or very near, the stellar surface. Coupled to the fact that in some stars, the absorption in DACs can almost saturate the line profile, 
this suggests that the surface perturbations causing them
can occupy a significant portion of the stellar disk (e.g., \citealt{1997A&A...327..699F, 2015ApJ...809...12M}).

While most of the aforementioned studies have stemmed from the observation of variability in the UV spectra of hot massive stars, there is an increasing
number of observational diagnostics that possibly reveal the presence 
of co-rotating bright spots on hot star surfaces, as well as the extended wind structures
which should result from their presence. For instance, DACs have been shown to be associated with 
variability in H$\alpha$ \citep{1997A&A...327..281K}.% as well as variability in excited state lines
%\citep{2015ApJ...809...12M}. 

While the CO96 model has been generally considered to successfully account for this phenomenon for the past 20 years, 
it has not since been revisited to include newly-derived
observational constraints\footnote{Some more sophisticated hydrodynamical simulations have been performed since the original CO96 study. 
They are carried out in three dimensions, and include slightly more detailed physics (e.g., \citealt{2004A&A...423..693D}, which notably concluded that
the CO96 2D approach was valid to derive the effect of the wind structures on the UV line profiles). However, these studies still used unrealistic spot
parameters with respect to the constraints described in the following paragraph.}. Photometric signatures related to putative spots generating CIRs 
have been claimed to have been found in a number of WR stars, e.g., WR110 \citep{2011ApJ...735...34C} and WR113 \citep{2012MNRAS.426.1720D}.
Typically, they involve light curves with seemingly stochastic variations, but time-frequency analysis reveals the presence of multiples of a
frequency the authors associate with rotation, appearing and disappearing depending on the number of structures in the wind at any given time.

\citet{2014MNRAS.441..910R} claimed the first detection of photometric variations due to co-rotating bright spots on the surface of an OB star
using broadband optical photometry from the \textit{MOST} space telescope. The star in question,
$\xi$ Persei, is an O7.5III(n)((f)) star which was observed by \textit{IUE} 
for 5 runs between 1987 and 1994 and shown to have well-defined DACs. The most important constraint derived in
the study of $\xi$ Persei's light curve is that the maximum peak-to-peak amplitude of the variations produced by the putative bright 
spots is about 10 mmag, or about 1\% of the apparent brightness. This sets
important limits on both the size and brightness contrast of the spots. 

In this study we will focus mostly on the classical DACs seen in UV resonance lines, as well as broadband optical photometry, and in particular on the possibility
of reconciling these various observations for a given star.  
The main CO96 model invoked spots with a 20\degree~ angular radius and a Gaussian 
brightness enhancement which peaked at a 50\% central brightness contrast relative to the surrounding photosphere. 
Such spots would generate photometric variations with an amplitude roughly 3 times larger than that observed in the light curve of $\xi$ Persei.

The goal of this paper is to couple the recent (optical) photometric and (UV) spectroscopic observations using a CO96-type phenomenological
model. Specifically, using the constraints derived photometrically to choose our input parameters, we aim to determine whether it is possible to reproduce, at least
qualitatively, the behaviour exhibited by the UV resonance lines of $\xi$ Persei. The numerical methods used to generate both the hydrodynamical wind simulations
and to compute synthetic line profiles are detailed in Section~\ref{sec:numerical}. In Section~\ref{sec:results}, we then describe the obtained results and 
compare them to the observational diagnostics. Finally, in Section~\ref{sec:concl} we draw conclusions and indicate the next steps toward a better understanding
of this phenomenon.

\section{Numerical methods}\label{sec:numerical}

\subsection{Hydrodynamical wind modelling}

The O star's wind is modelled in 2D in the equatorial plane using VH-1, a multidimensional ideal compressible hydrodynamics code written in FORTRAN which
uses the piecewise parabolic method (PPM) algorithm developed by \citet{1984JCoPh..54..174C}.
However, the key factor in forming CIRs is the variation in the line-driving force due to inhomogeneities on the surface of the star. \citet{1995ApJ...440..308C} have 
shown how to compute the vector line force for such a flux distribution (specifically in the context of an oblate finite disk; OFD). Therefore, using a 
FORTRAN subroutine (\textit{gcak3d}) that implements this method together with VH-1, we can perform a full radiation hydrodynamics simulation of the 
wind\footnote{Note that, for the calculations presented in this paper, we only compute the radial component of the line-driving force.}.
However, rather than implementing an OFD, we assume a spherical star and 
implement Gaussian bright spots on the equator. Their ``amplitude" ($A$) corresponds to the maximum flux enhancement
at the peak of the distribution
(the difference between the flux at the center of the spot and the unperturbed flux, divided by the latter), 
and their ``angular radius" ($r$) corresponds to the standard deviation of the Gaussian distribution multiplied
by a factor of $\sqrt 2$. Using both of these spot parameters $A$ and $r$, 
we can infer the fractional amplitude of the variation ($A_{\textrm{var}}$) a single spot would cause in the disk-integrated light curve of the star at a
given wavelength or within a given bandpass. 

Using $\theta$ as the angle between the surface normal and the observer's line of sight at a given point on the stellar surface (or in other words the limb angle)
and considering the maximum flux due to the addition of a Gaussian spot in the center of the disk, we can calculate $A_{\textrm{var}}$ as
the ratio between the additional intensity due to the spot and the integrated intensity of the star without the spot. We can express this quantity
as a function of the amplitude and radius of the spot using the following equation:

\begin{equation}
\frac{\int_{0}^{\pi/2} A \textrm{e}^{- \theta^2/r^2} \sin \theta \cos \theta d\theta}{\int_{0}^{\pi/2} \sin \theta \cos \theta d\theta} = A_{\textrm{var}} \approx A \sin^2 r\,,
\end{equation}

%\noindent where $A_{\textrm{var}}$ corresponds to the fractional amplitude of the variations in the light curve. 
\noindent where $A$ and $A_{\textrm{var}}$ are dimensionless (they correspond to ratios) and $\theta$ is in radians. For small values of $r$, this reduces to

\begin{equation}\label{eq:spotconstr}
A_{\textrm{var}} \approx A r^2\,.
\end{equation}

\begin{figure}
\begin{center}
%\vspace{3.5cm}
\includegraphics[width=3.0in]{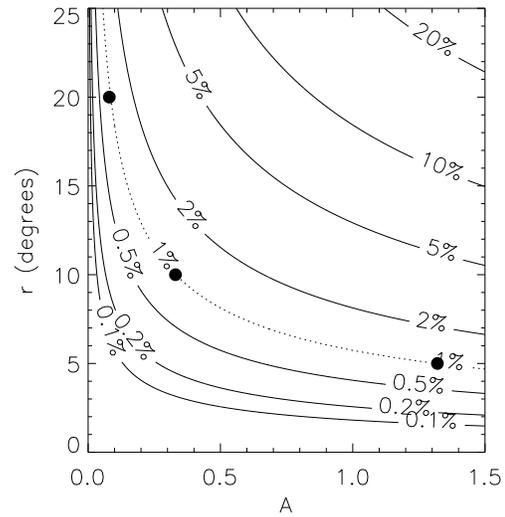}
\caption{Amplitude of the optical photometric variations ($A_{\textrm{var}}$) 
as a function of the spot parameters $A$ and $r$. The dotted line represents the maximum
amplitude of variability seen in the light curve of $\xi$ Persei; the large black dots represent the three sets of model parameters used later in this study, as
summarized in Table~\ref{tab:models}.}
\label{fig:avar}
\end{center}
\end{figure}

\noindent The behaviour of $A_{\textrm{var}}$ as a function of $A$ and $r$ is shown in Fig.~\ref{fig:avar}. Another important consideration in our case
is the fact that there is no reason, in general, to expect the flux contrast in the bright spots to be the same at all wavelengths. In particular, if we consider
both the spots and the stellar surface to have a black-body spectral energy distribution, then we can estimate the different enhancements at various
wavelengths. While bright spots have been inferred to exist on $\xi$ Persei's surface using optical photometry, the relevant wavelength regime in terms
of wind-driving is in the ultraviolet. Using Eq.~\ref{eq:spotconstr}, we can choose
pairs of values of spot size and amplitude that correspond to the 10 mmag photometric amplitude 
found by \citet{2014MNRAS.441..910R} in the optical (or in other words, for which $A_{\textrm{var}} = 0.01$). We can then estimate the 
associated flux enhancement in the ultraviolet by finding the temperature $T'$ at which:

\begin{equation}\label{eq:planck_spot}
\frac{B(\lambda_{\textrm{opt}}, T')}{B(\lambda_{\textrm{opt}}, T_{\textrm{eff}})} - 1 = A_{\textrm{opt}}
\end{equation}

\noindent where $B$ is the Planck function, $\lambda_{\textrm{opt}}$ is a wavelength representative of the optical bandpass (here we choose 5000 \AA) and
$A_{\textrm{opt}}$ corresponds to the optical brightness enhancement of the spot. We then compute the UV enhancement in a similar fashion:

\begin{equation}\label{eq:planck_spot}
\frac{B(\lambda_{\textrm{UV}}, T')}{B(\lambda_{\textrm{UV}}, T_{\textrm{eff}})} - 1 = A_{\textrm{UV}}
\end{equation}
  
\noindent where $\lambda_{\textrm{UV}}$ is chosen to be 1500 \AA~ and $A_{\textrm{UV}}$ is the UV brightness contrast of the spots, which will be used
in the hydrodynamical models. We select three models respecting these constraints, as summarized in
Table~\ref{tab:models} and illustrated in Fig.~\ref{fig:avar}. 

We let the spot radius vary between 5\degree and 20\degree, a range which is justified by the fact that (i) 
the spots must not be too
small since they must cover a significant fraction of the stellar disk, and (ii) they also cannot be too large, otherwise
the corresponding brightness contrast might be too low to produce noticeable perturbations in the wind (this point will be further
discussed in Section~\ref{sec:concl}).

Another important change in this study, as compared to CO96, is the implementation of the 
ionization parameter $\delta$, as derived by \citet{1982ApJ...259..282A}. This
allows us to take into account the radial dependence of the ionization throughout the wind, 
which in turn influences the local electron density and therefore the line driving.
A higher value of $\delta$ leads to a greater driving force in the inner wind, ultimately ejecting too much material from the surface for the
radiation to be able to continuously accelerate it outwards, or in other words ``overloading" the wind. \citet{1982ApJ...259..282A} finds a typical value
of $\delta \approx 0.1$ for massive O stars. Including this factor might allow weaker spots to form CIRs. 
Therefore each of the three spot models described above is also divided into two sub-models: one that does not take the ionization factor into account, and one that
does (see Table~\ref{tab:models} for a summary of the 6 models). A final departure from the method used in CO96 is that rather than using a heuristic
scaling formula for enhanced line-driving, we use multiple ray quadrature to compute the line-force from a localized bright spot.

\begin{table}
\begin{center}
\caption[Model list]{List of models used in this study; $r$ corresponds to the angular radius of the spot, $A$ is the central amplitude of the spot
flux enhancement and $\delta$ is the ionization parameter used to calculate the line-driving force.}\label{tab:models}
\begin{tabular}{|l|r|c|c|c|l|}
  \hline
Model & $r$ & $A_{\textrm{opt}}$ & $T'$ & $A_{\textrm{UV}}$ & $\delta$\\
   & (\degree) &  & (kK) &  &   \\
  \hline
1A  & 5  & 1.32 & 67.9 & 3.30 & 0 \\
1B  & 5  & 1.32 & 67.9 & 3.30 & 0.1 \\
2A  & 10 & 0.33 & 44.1 & 0.71 & 0 \\
2B  & 10 & 0.33 & 44.1 & 0.71 & 0.1 \\
3A  & 20 & 0.08 & 38.0 & 0.16 & 0 \\
3B  & 20 & 0.08 & 38.0 & 0.16 & 0.1 \\
  \hline
\end{tabular}
\end{center}
\end{table}

We model a 90 degree sector of the equatorial plane, using 90 azimuthal zones and 250 radial zones. The radial zones are spaced geometrically, each zone being 2\%
larger than the previous one, and they span a region extending from the stellar surface ($R = R_{*}$) to 10 stellar radii ($R = 10~ R_{*}$). 
The azimuthal boundary conditions are periodic, such that we are actually modelling 4 identical and 
equally spaced equatorial bright spots\footnote{Of course, such a distribution would lead to somewhat smaller photometric variations than those caused
by a single spot, since there would always
be at least one spot visible on the stellar surface. This means that in general, due to averaging effects, we could have used stronger and/or bigger spots
and still respect the 10 mmag constraint; therefore, the spot parameters we are using for our various models correspond to conservative estimates.}.
In order to properly resolve the surface and to account for the additional driving provided by even the smallest spots, we set up a numerical
quadrature, with rays intersecting the star at various values of $p$ (impact parameter) and $\phi'$ (azimuthal angle).
These points on the star are distributed using a Gauss-Legendre quadrature in $p^2$ and $\phi'$, and using a rotation factor which varies the $\phi'$
values between different values of $p$ to better probe the stellar disk (as explored by \citealt{2015PhDT..........K}).
First we run a 1D simulation for 600 ks to obtain a relaxed, spherically-symmetric wind that behaves like a typical line-driven wind, as described 
by \citet{1975ApJ...195..157C}, henceforth referred to as ``CAK theory". We then use this as input for a 2D simulation, also with
a uniform surface flux distribution, and relax it for 700 ks. Finally, 
we ``turn on" the spots and let the simulation run for 800 ks (until it reaches a steady-state solution). We compute 14 snapshots
which are 5~ks apart (they span a quarter of the rotational period) to trace the temporal variation of the wind. 
All models use the same input stellar and modelling parameters, detailed in Table~\ref{tab:param}.

\begin{table}
\begin{center}
\caption[Model parameters]{Parameters used to generate the hydrodynamical models.}\label{tab:param}
\begin{tabular}{|l|c|}
  \hline
Model parameter & Value\\
  \hline
Stellar mass $M_{*}$                     & $26~ M_{\odot}$                 \\
Stellar luminosity $L_{*}$               & $2.6 \times 10^5 L_{\odot}$    \\
Stellar radius $R_{*}$                   & $14~ R_{\odot}$                 \\
Effective temperature $T_{\textrm{eff}}$ & 36.0 kK                        \\
Surface azimuthal velocity $v_{\phi, 0}$ & $2.2 \times 10^7$ cm/s         \\
Surface density $\rho_{0}$               & $3.0 \times 10^{-11}$ g/cm$^3$ \\
CAK power-law index $\alpha$             & 0.6                            \\
Collective line force $\bar{Q}$          & $10^3$                         \\
Quadrature points in $p$ and $\phi'$     & $9 \times 9$                   \\
$p$ rotation factor                      & 0.33                           \\
  \hline
\end{tabular}
\end{center}
\end{table}

The stellar parameters for $\xi$ Persei are obtained from \citet{2004A&A...415..349R} ($M_{*}$, $L_{*}$), \citet{2010A&A...519A..50K} ($R_{*}$)
and \citet{2014MNRAS.444..429D} ($T_{\textrm{eff}}$). 
The surface azimuthal velocity is chosen to be equal to the
value of the projected rotational velocity ($v \sin i$) reported by \citet{2014MNRAS.444..429D}. The surface density is adjusted to ensure that the wind outflow
is initially subsonic as it leaves the stellar surface (otherwise, that would lead to significant instabilities in the simulation). 
We chose a standard value of 0.6 for the CAK power-law index (e.g., \citealt{2000A&AS..141...23P}). The line force normalization, $\bar{Q}$, 
is an important input parameter as it determines the global behaviour of 
the wind by calibrating the force of the line-driving mechanism \citep{1995ApJ...454..410G}. A typical value for 
OB stars was chosen for this parameter, although its exact value is not relevant, since for this study we are more interested in the structures
that form in the wind than in the overall wind properties.

Finally, the code outputs a 2D grid mapping of the density, radial velocity and azimuthal velocity of the wind in the equatorial plane.

\subsection{Line profile synthesis}

Once the 2D wind models are generated, they must first be extrapolated into a three-dimensional grid in order to perform the line synthesis. We use the
same prescription as CO96, as described in their Eq. 21, generating 181 latitudinal zones to ensure that grid cells near the equator have 
comparable sizes in the latitudinal and azimuthal directions. The value of the latitudinal spread parameter (corresponding to their ``$\sigma$") 
used to perform that extrapolation
for a given model corresponds to the angular radius of the spot used in that model.
We then use Sobolev Exact Integration (SEI, \citealt{1987ApJ...314..726L}) to compute the line profile. Our calculation is based on the ``3D-SEI" method 
(CO96) and
uses the same code as \citet{2013MNRAS.431.2253M}. This code solves the formal integral of radiative transfer in a 3D cylindrical coordinate system
aligned toward the observer (following \citealt{2012MNRAS.423L..21S}).

A few important input parameters are used. First, the spectral resolution depends on the number of velocity bins computed across the entire line profile.
We use 161 points for all models, which leads roughly to a 45 km/s resolution for the models without the ionization factor, and a 30 km/s resolution
for the models which include the ionization factor. %The latter case corresponds to a spectral resolution of $R = c/\Delta v$ of around $10^{4}$, which
%corresponds to the resolution obtained using the Short Wavelength Prime (SWP) camera on \textit{IUE} \citep{1996A&AS..116..257K}, therefore facilitating the direct %comparison
%of our synthetic line profiles to the observed behaviours. 
We also use a high number of rays (318 rings of 401 azimuthal rays in the cylindrical coordinate system) 
to perform the formal integral and compute emergent flux profiles so as to probe our grid as finely
as possible and yield the most precise results. As for the line strength, we use the $\kappa_{0}$ parameter as defined in Eq. 13 of 
\citet{2014A&A...568A..59S}. This dimensionless parameter, first developed by \citet{1981A&A....93..353H} and then extended to be applicable to
full radiation-hydrodynamical simulations of unstable winds (e.g., \citealt{1993A&A...279..457P, 2010A&A...510A..11S}),
determines the optical depth and is proportional to the mass-loss rate and to the fractional abundance of the absorbing ion for a given line. 
We chose a value of 1.0, which corresponds to a marginally optically thick, but unsaturated line analogous
to the resonance lines used in typical DAC studies (such as the Si \textsc{iv} doublet). We also did not use an underlying photospheric profile as a
lower boundary condition.

\begin{figure}
\begin{center}
%\vspace{3.5cm}
\includegraphics[width=3.0in]{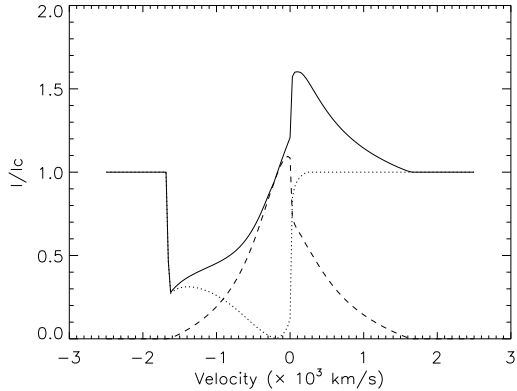}
\caption{Computed P Cygni line profile for a line strength of $\kappa_{0} = 1.0$ using an 
unperturbed wind model (with $\delta$ = 0.1). The dotted line shows the absorption component, while the dashed line shows the emission component;
the full line corresponds to the total profile. We can see that the absorption is already very strong at around $v/v_{\infty} = 0.3$ (which corresponds to about
500 km/s for this profile).}
\label{fig:abs}
\end{center}
\end{figure}

Finally, the non-monotonic velocities induced by the Line-Deshadowing Instability \citep{1988ApJ...335..914O} can be modelled approximately in these 
line-profile calculations as a turbulent
velocity parameter $v_{\textrm{turb}}$ 
which is implemented like microturbulence and made to vary between the base of the wind and its outer regions (e.g., \citealt{1995A&A...295..136H}).
The inclusion of such a parameter was tested, but led to features that were qualitatively different from the observed DACs, so $v_{\textrm{turb}}$ is 
set to zero\footnote{Such a treatment is reasonable since it has already been shown that CIR-like density structures can inhibit the growth
of instabilities, thus greatly decreasing this velocity dispersion \citep{1999LNP...523..294O}. Presumably, to reproduce the shape of the line's blue edge,
instability should be taken into account in the ``unperturbed" wind regions between the CIRs, but since it does not constitute the focus of this study,
this was not implemented in the line transfer code.}.
An example of the resulting profile, including the individual absorption and emission contributions, is calculated in the case of an
unperturbed, spherically symmetric wind (modelled using the parameters in Table~\ref{tab:param}) and shown in Fig.~\ref{fig:abs}.

\section{Results}\label{sec:results}

Each individual step in the calculations allows us to draw a number of conclusions 
about the surface spots and their effect on the wind. Therefore we present the hydrodynamical
simulations and the associated computed resonance line profiles separately in the following subsections.

\subsection{Wind properties}

We first investigate the relative change in the global parameters of the wind between the various models summarised in Table~\ref{tab:models}. 
The global mass-loss rate can be calculated by computing:

\begin{equation}
\dot{M} = 4 \pi R^2 \rho v
\end{equation}

\noindent at a given radial distance $R$ from the centre of the star. 
For an unperturbed wind, $\rho$ and $v$ should be independent of the azimuthal angle $\phi$, but that is not the case once the spots
are introduced. We can therefore compute for each model an unperturbed mass-loss rate, as well as the mass-loss rate including the effect of the spots to
determine the amount of extra material ejected by the spots. In the latter case, we average $\rho$ and $v$ azimuthally. 

Similarly, we can approximate the terminal
velocity $v_{\infty}$ by taking the maximum radial velocity $v_{\textrm{max}}$ 
within the simulation zone, and examine how it varies between the unperturbed winds and the spotty models.
Table~\ref{tab:wprop} summarizes these wind properties, as well as other results from the next subsection.

\begin{table*}
\begin{center}
\caption[Model properties]{Observed properties for each of our models. The first four variables refer to wind properties. $\dot{M}_{0}$
is the unperturbed mass-loss rate, $\dot{M}_{\textrm{s}}$ is the ``spotty" mass-loss rate, $v_{\textrm{max},0}$ is the unperturbed terminal velocity and
$v_{\textrm{max, s}}$ is the ``spotty" terminal velocity. The last two variables are related to the synthetic line profiles: $v_{\textrm{start}}/v_{\infty}$
corresponds to the approximate fraction of the terminal velocity at which the DACs appear, and $F_{\textrm{min}}/F_{\textrm{c}}$ is the maximum
DAC depth (expressed as the minimum value found in the
quotient spectra).}\label{tab:wprop}
\begin{tabular}{|l|c|c|c|c|c|c|}
  \hline
Model & $\dot{M}_{0}$ & $\dot{M}_{\textrm{s}}$ & $v_{\textrm{max},0}$ & $v_{\textrm{max, s}}$ & $v_{\textrm{start}}/v_{\infty}$ & $F_{\textrm{min}}/F_{\textrm{c}}$\\
   & ($10^{-7} M_{\odot}$/yr) & ($10^{-7} M_{\odot}$/yr) & ($10^{3}$ km/s) & ($10^{3}$ km/s) &  &  \\
  \hline
1A  & 6.41 &  10.1 & 2.33 & 4.01 & N/A  & 0.392\\
1B  & 8.56 &  15.2 & 1.66 & 2.68 & N/A  & 0.448\\
2A  & 6.41 &  8.09 & 2.33 & 2.78 & 0.41 & 0.317\\ 
2B  & 8.56 &  11.0 & 1.66 & 1.90 & 0.46 & 0.309\\
3A  & 6.41 &  7.10 & 2.33 & 2.43 & 0.47 & 0.262\\
3B  & 8.56 &  9.69 & 1.66 & 1.69 & 0.47 & 0.227\\
  \hline
\end{tabular}
\end{center}
\end{table*}

Examining Table~\ref{tab:wprop}, we first note that the mass-loss rates are of the same order of magnitude, although slightly larger, 
than some of the observationally-determined values in the literature 
(or than those predicted by the empirical relation derived by \citealt{1993ApJ...412..771L}). 
While it is known that theoretically-computed mass-loss rates often differ from empirical values by a factor of a few,
part of the discrepancy might also come from the choice of the line-force parameters $\bar{Q}$, $\alpha$ and $\delta$ 
(e.g., \citealt{1995ApJ...454..410G}). The inclusion of bright spots
increases the overall mass-loss rate, as more material is driven from the surface at their location.

Furthermore, we notice that using this set of line-force parameters,
the unperturbed terminal velocity for $\delta = 0$ corresponds exactly to the empirical value reported for $\xi$ Persei 
by \citet{1996A&AS..116..257K}. However,
when the effect of the ionization parameter is included, the terminal velocity decreases significantly. 
Ultimately, while these results inform us about the global wind properties, our goal is not to reproduce them. 
The terminal velocity and mass-loss rate also depend on other line-driving parameters,
not solely on the ionization parameter.
As a consequence, in Section~\ref{ssec:seiresults} we treat the terminal velocity as a scaling parameter and plot our dynamic spectra in terms of $v/v_{\infty}$.

\begin{figure*}
\begin{center}
%\vspace{3.5cm}
\subfigure[Model 1A]{\includegraphics[width=3.4in]{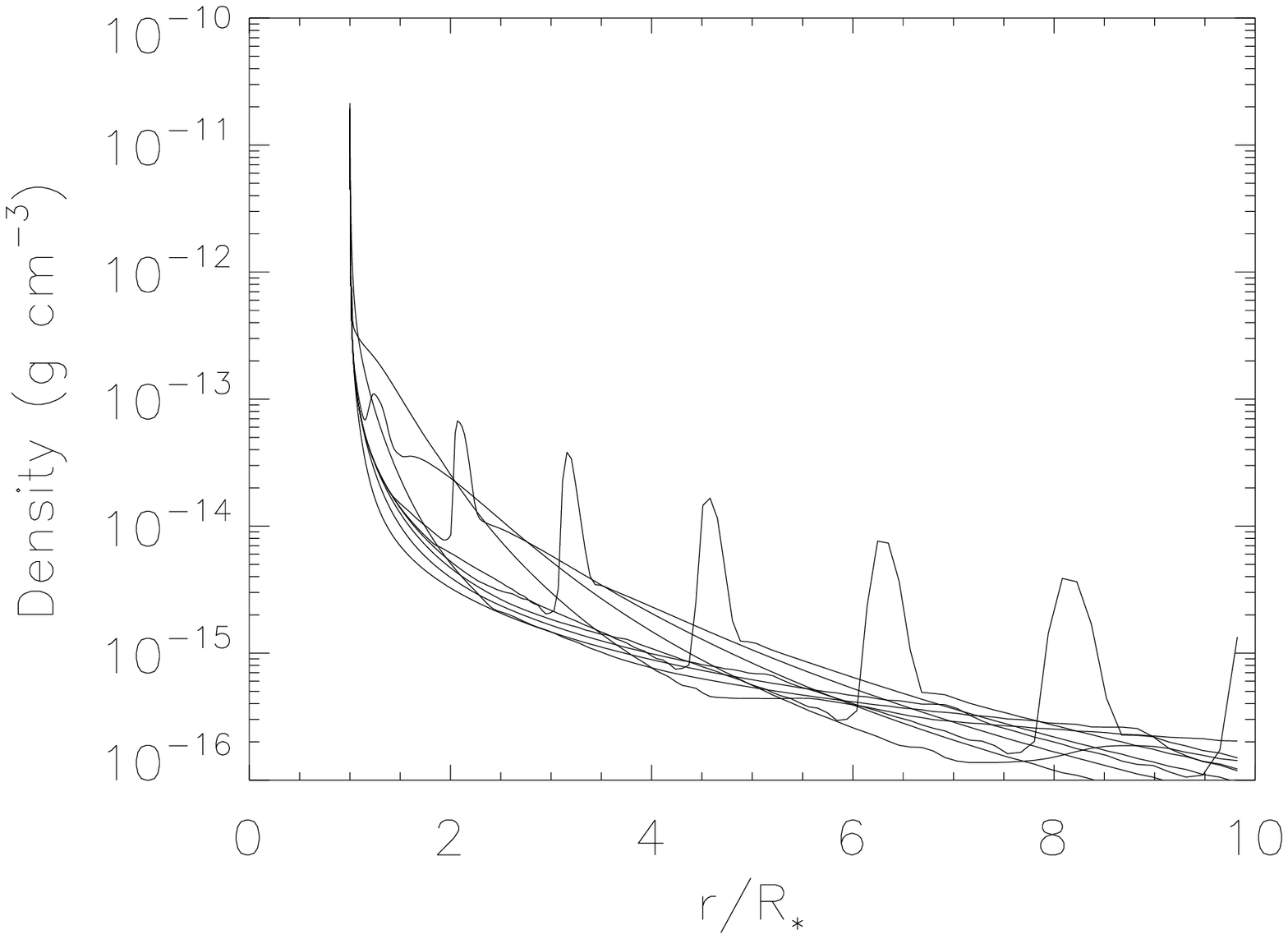}}
\subfigure[Model 1B]{\includegraphics[width=3.4in]{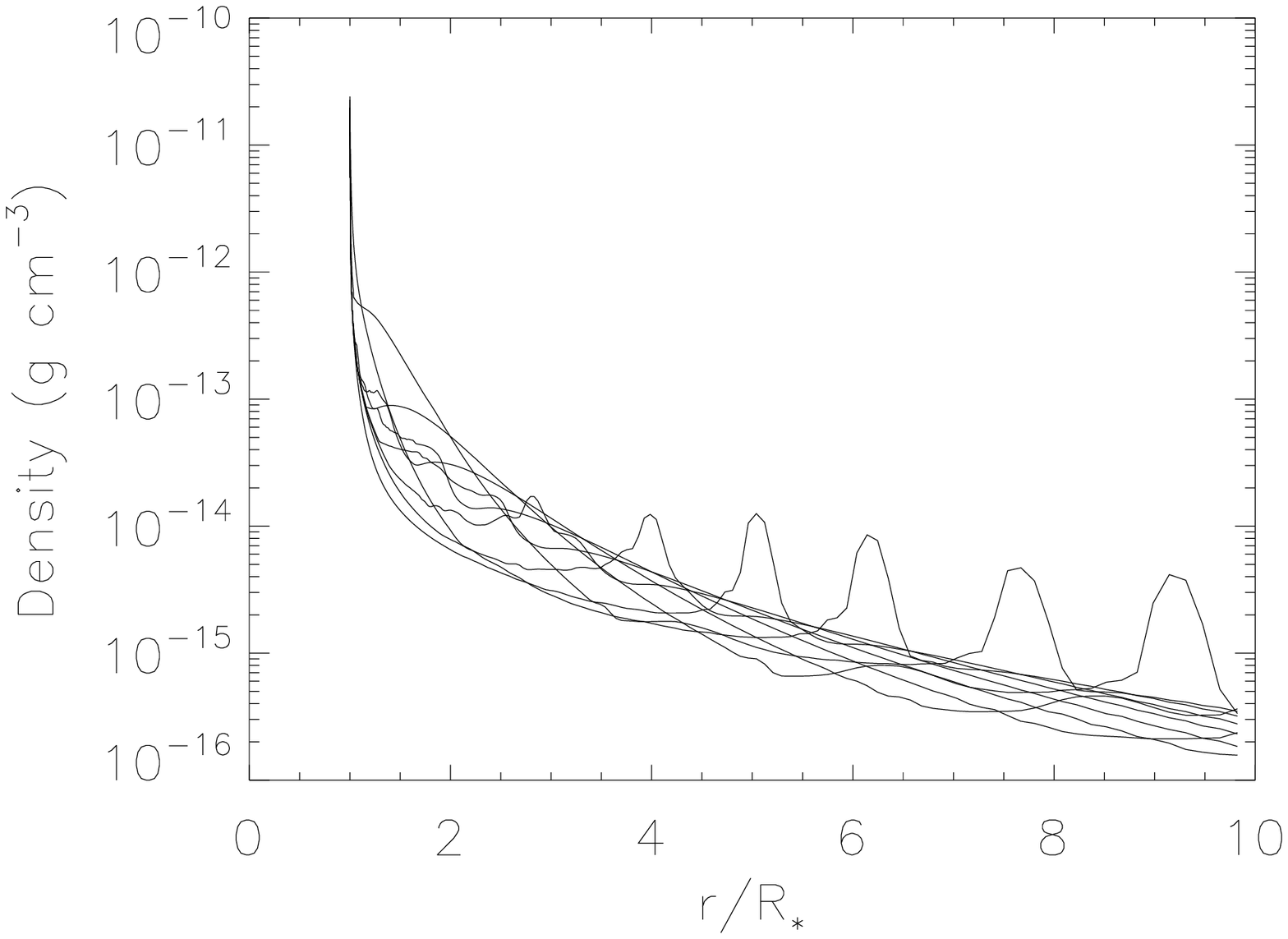}}
\subfigure[Model 2A]{\includegraphics[width=3.4in]{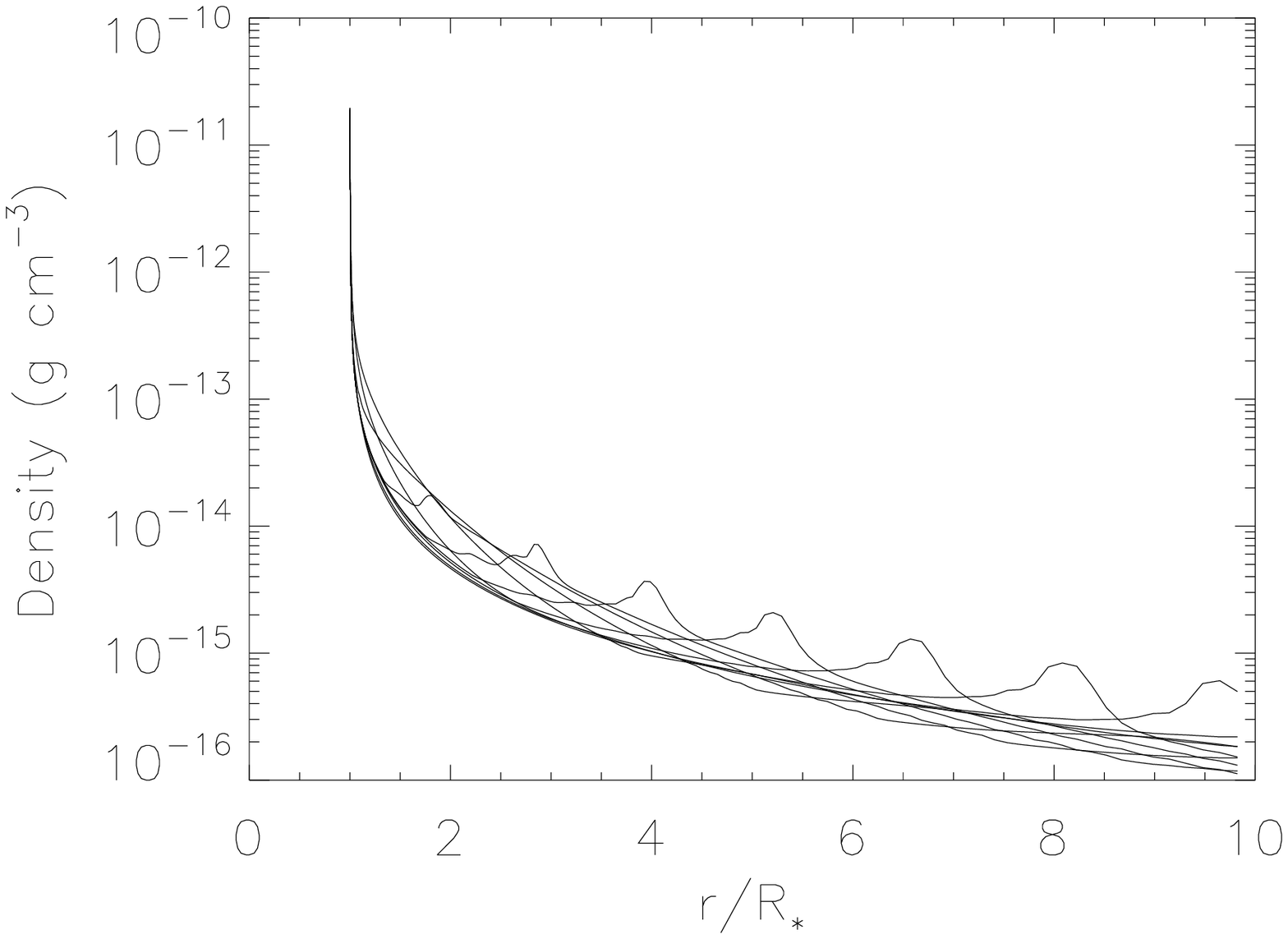}}
\subfigure[Model 2B]{\includegraphics[width=3.4in]{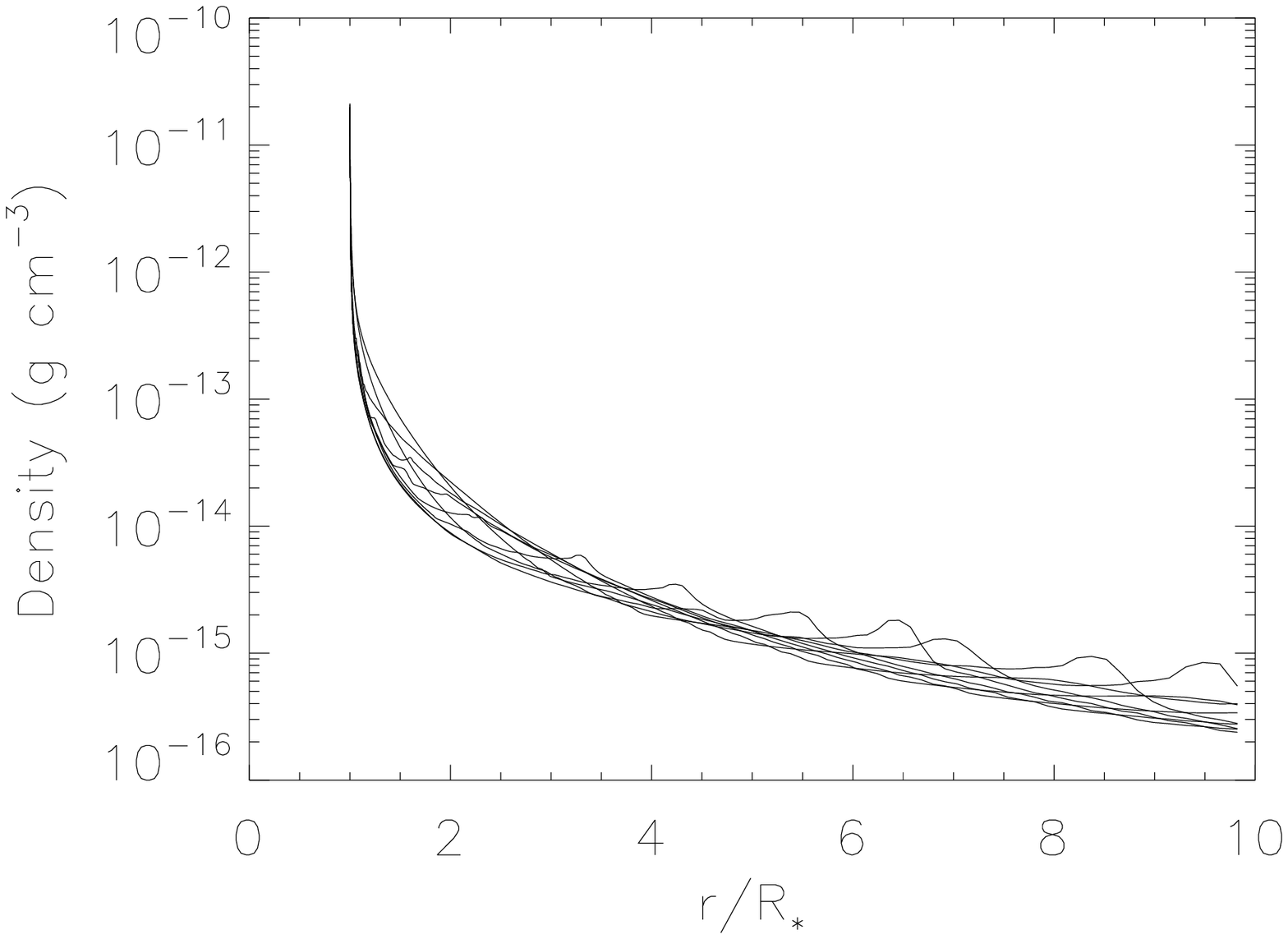}}
\subfigure[Model 3A]{\includegraphics[width=3.4in]{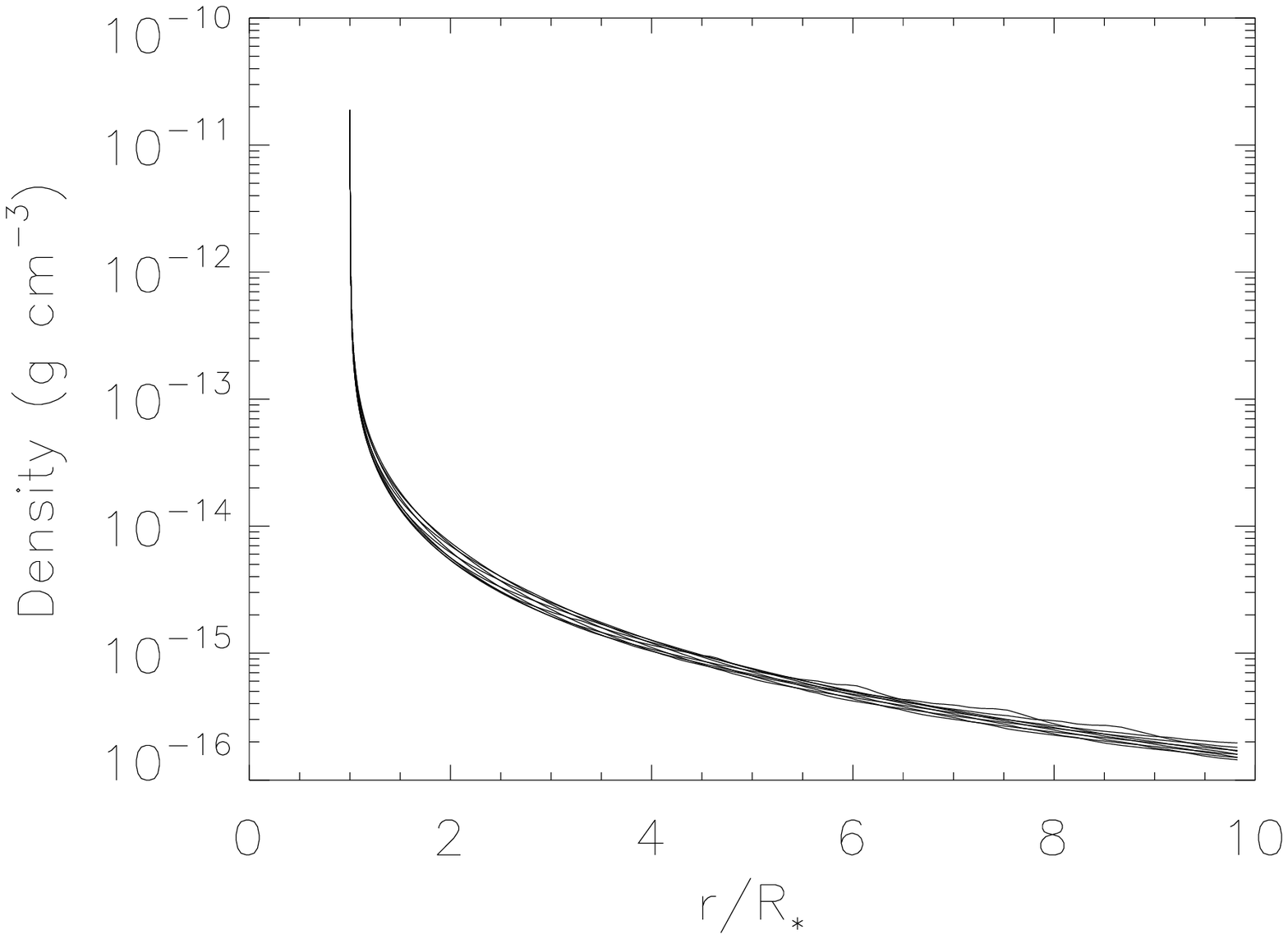}}
\subfigure[Model 3B]{\includegraphics[width=3.4in]{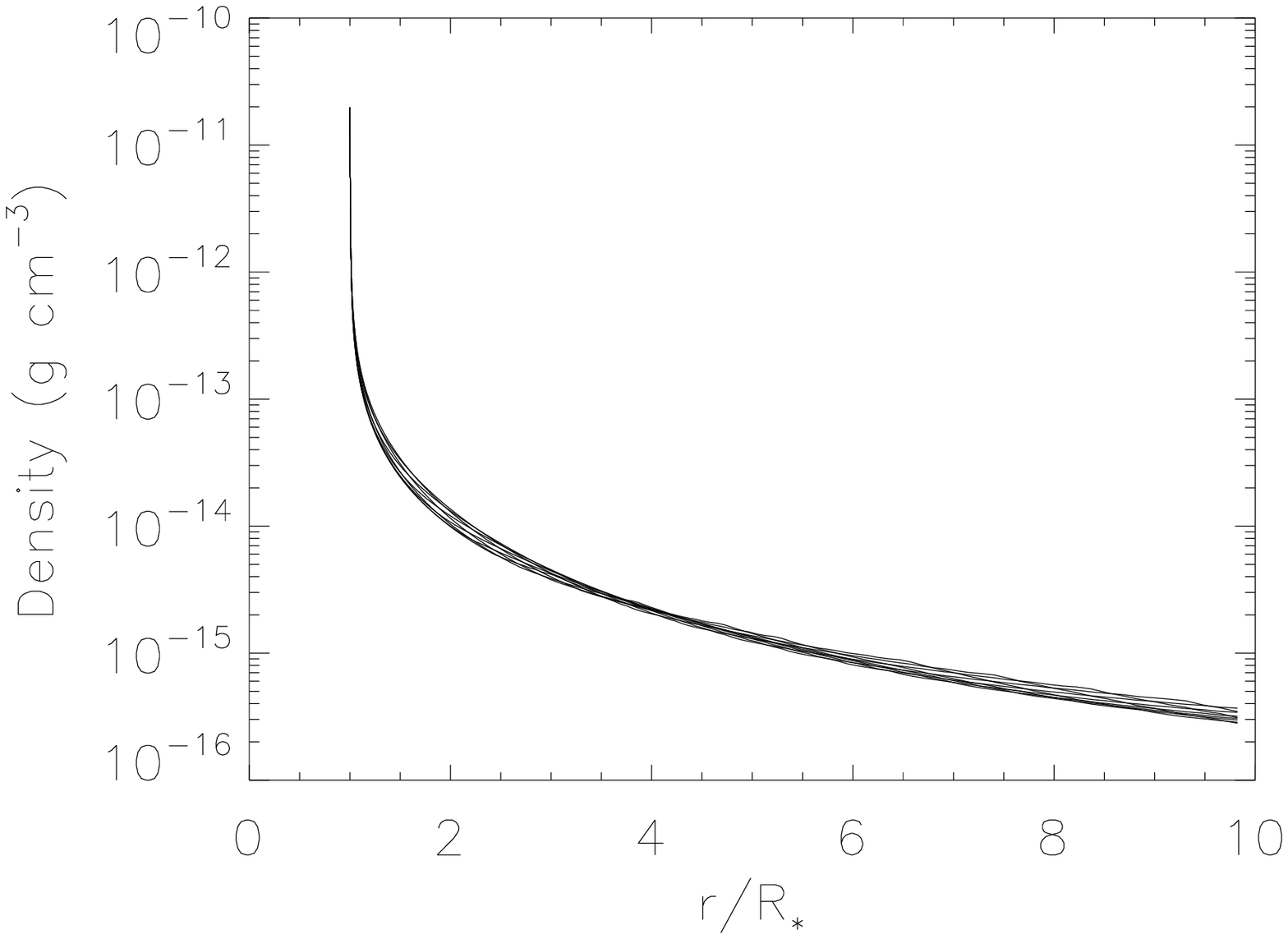}}
\caption{Radial dependence of the density for each of our 6 models, shown for azimuthal sectors which are 10\degree \ apart.}
\label{fig:rhor}
\end{center}
\end{figure*}

\begin{figure*}
\begin{center}
%\vspace{3.5cm}
\subfigure[Model 1A]{\includegraphics[width=3.4in]{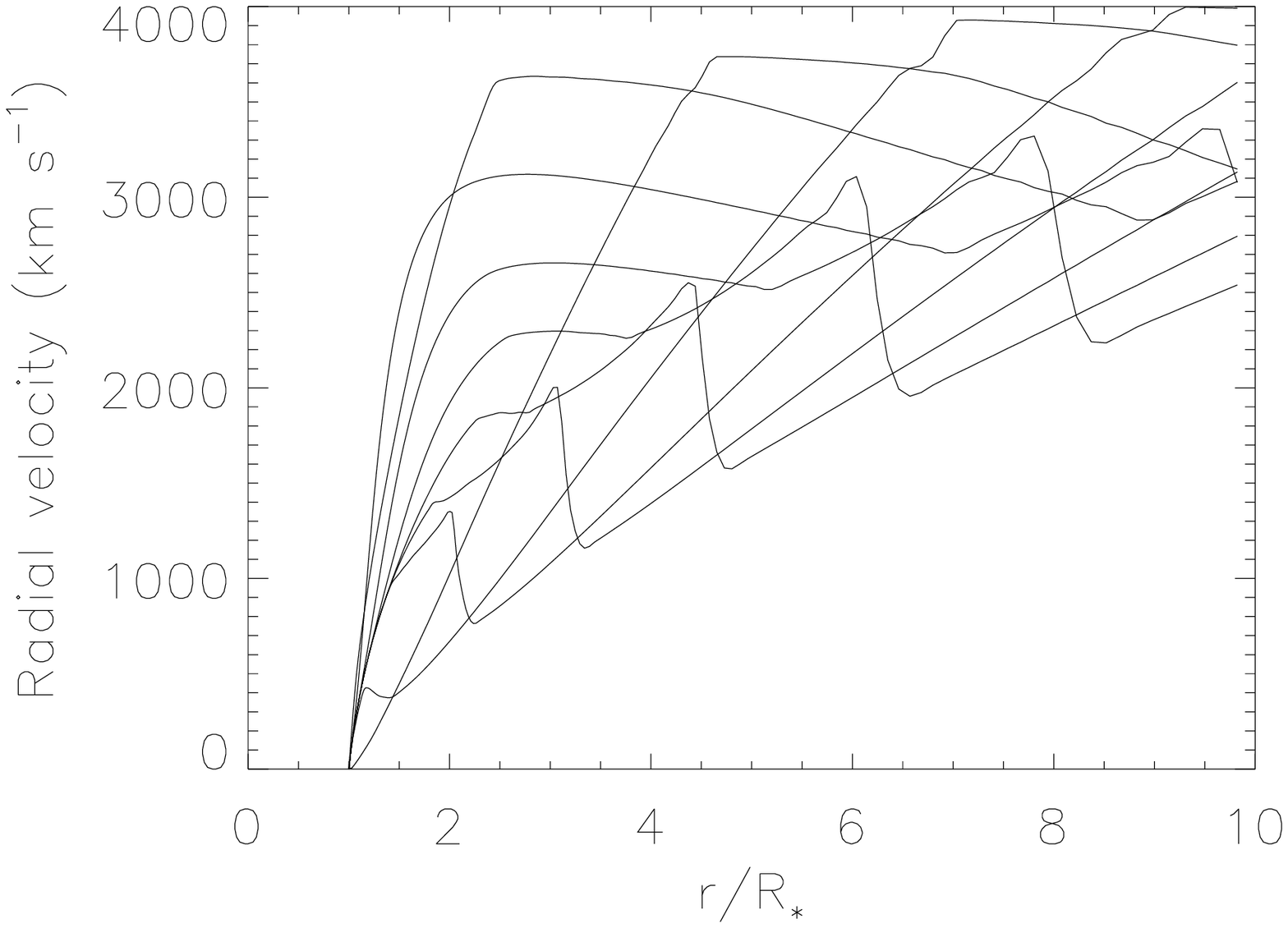}}
\subfigure[Model 1B]{\includegraphics[width=3.4in]{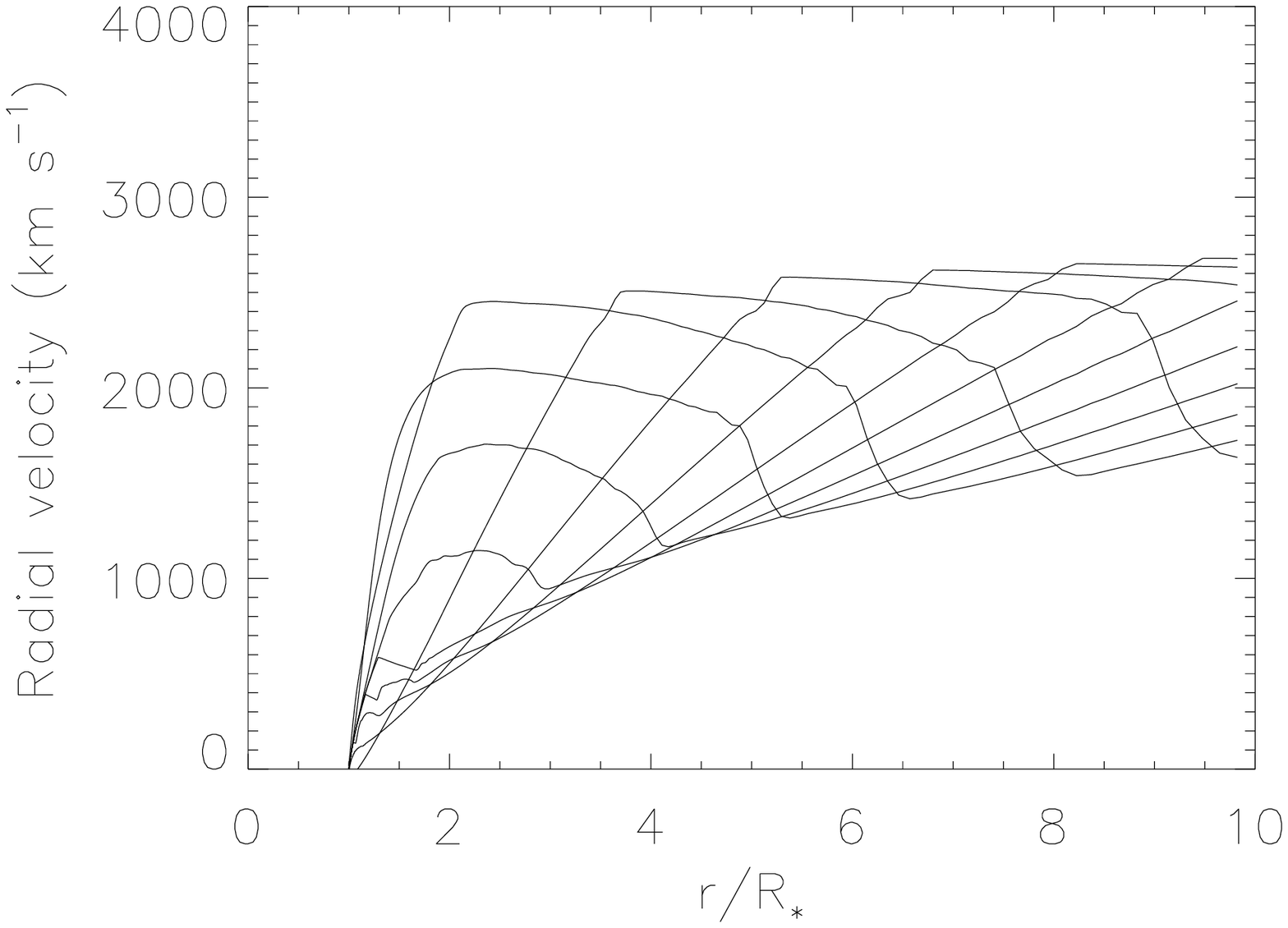}}
\subfigure[Model 2A]{\includegraphics[width=3.4in]{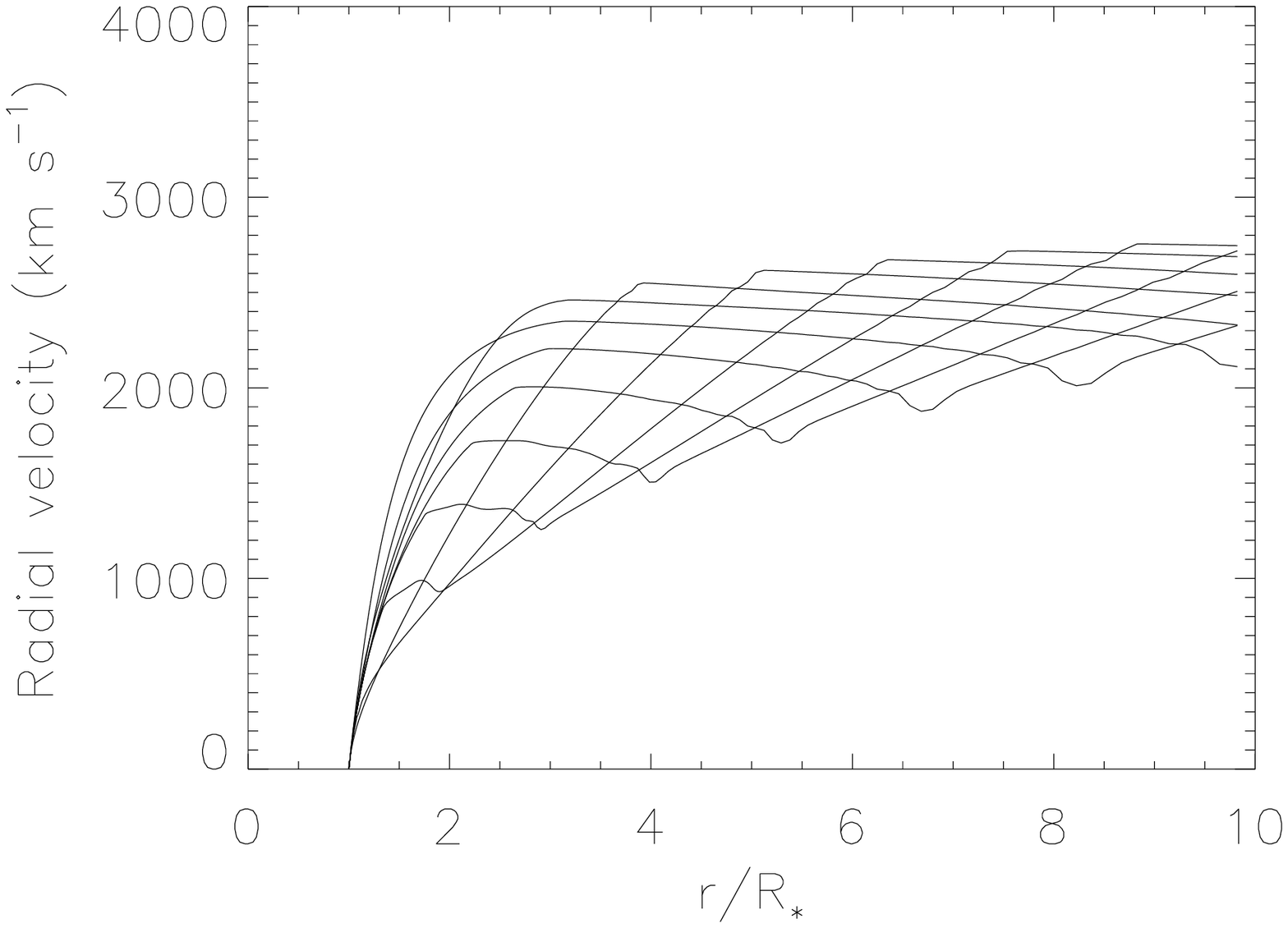}}
\subfigure[Model 2B]{\includegraphics[width=3.4in]{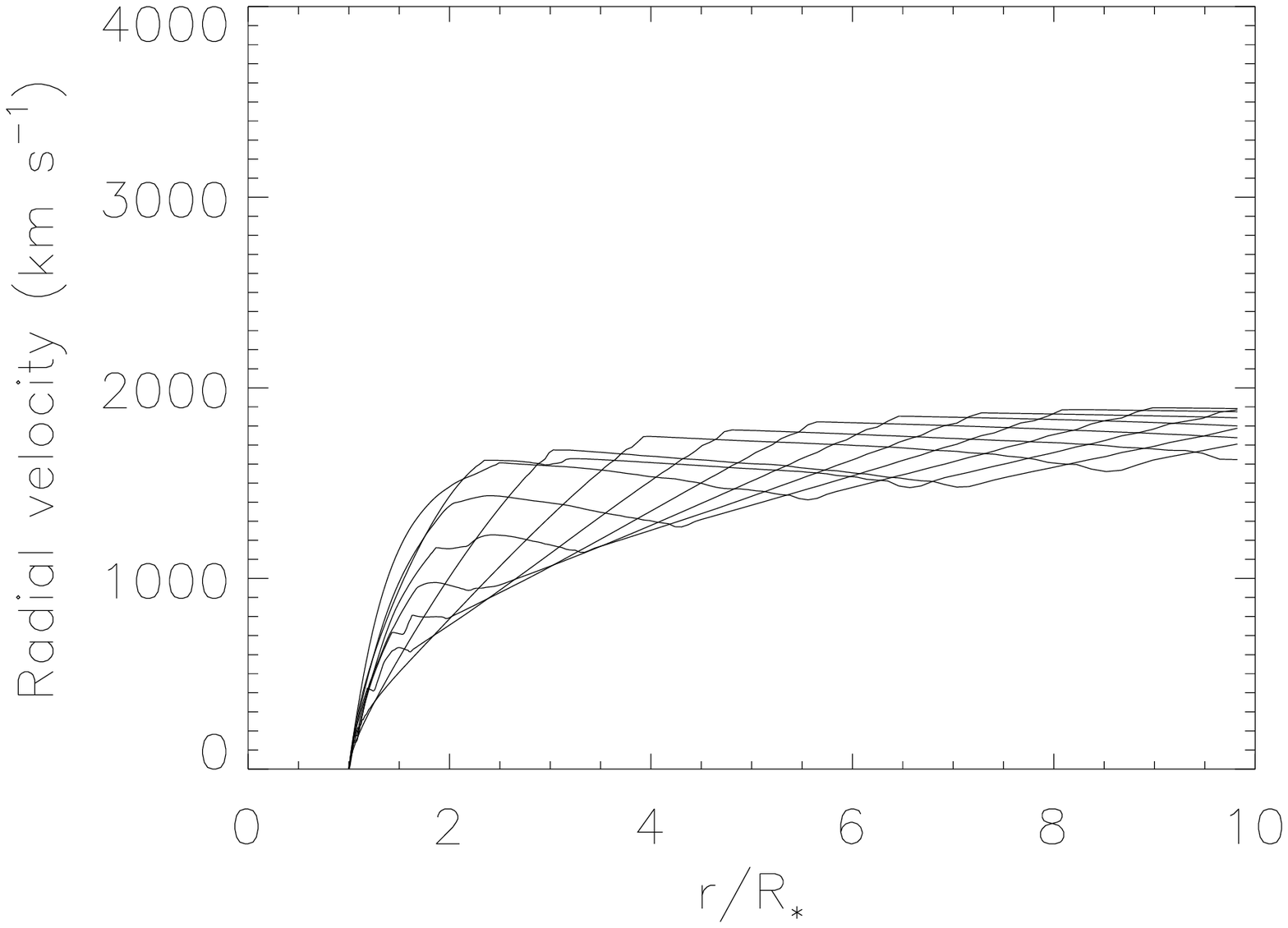}}
\subfigure[Model 3A]{\includegraphics[width=3.4in]{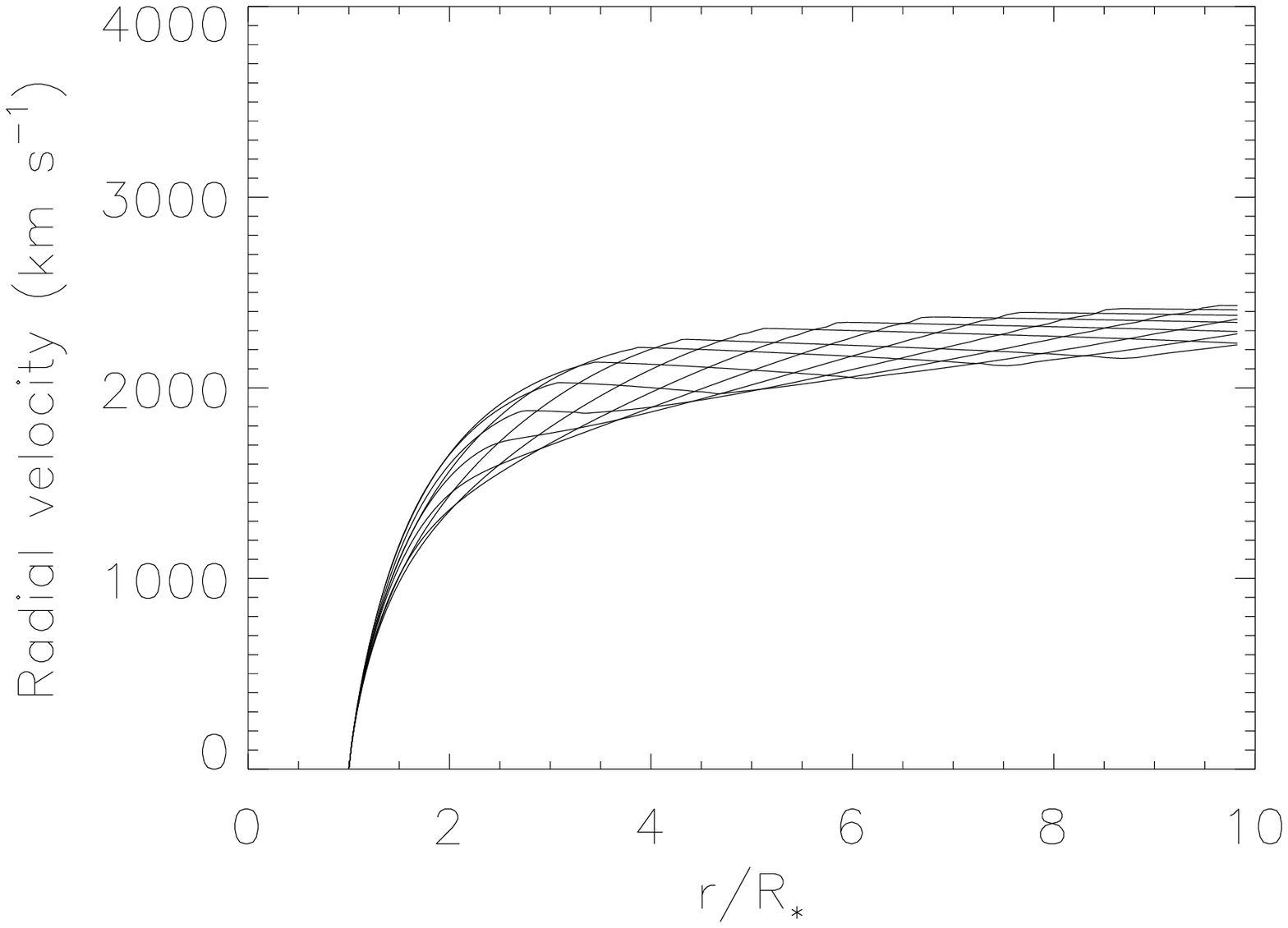}}
\subfigure[Model 3B]{\includegraphics[width=3.4in]{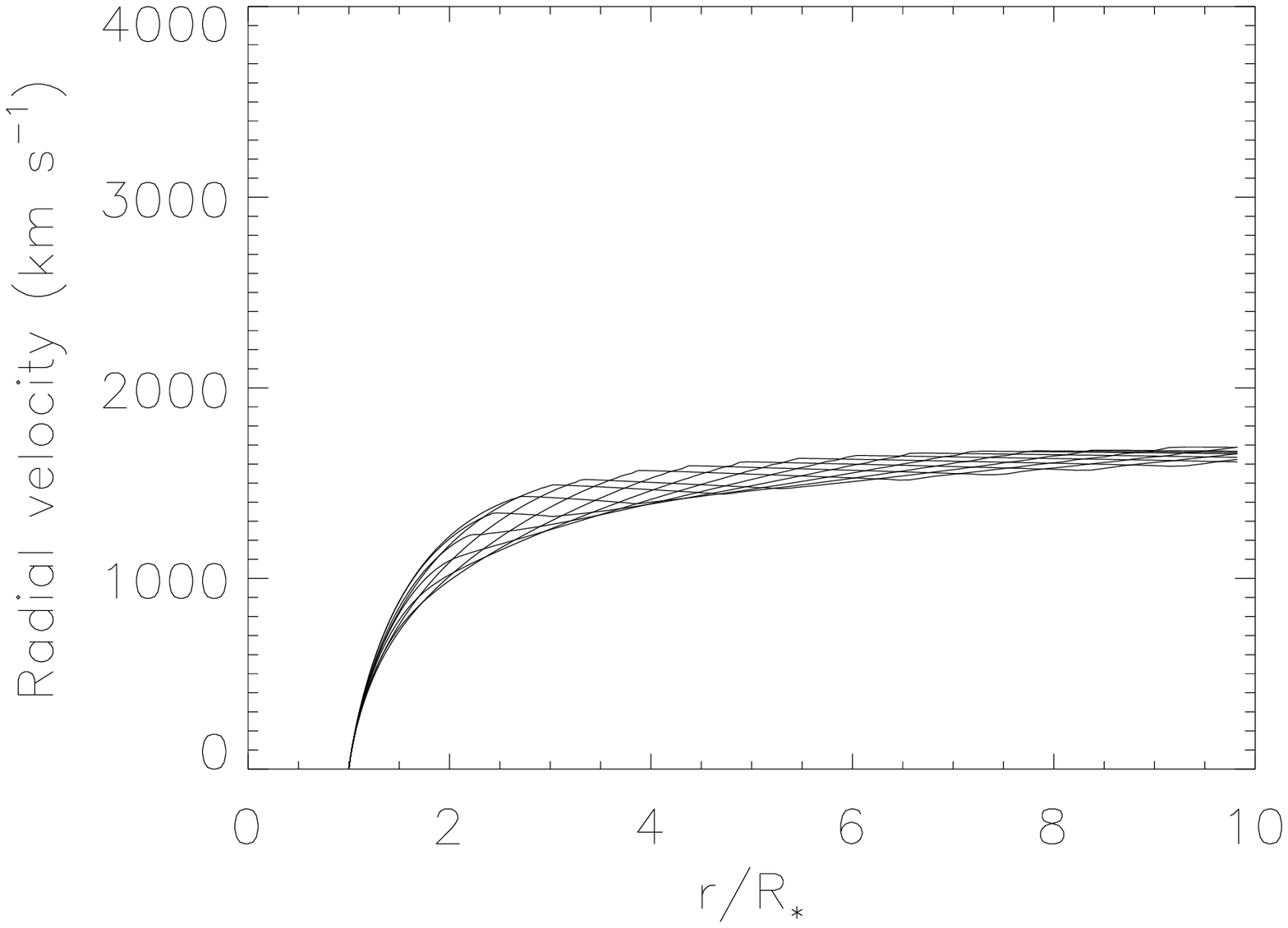}}
\caption{Same as Fig.~\ref{fig:rhor}, but for radial velocity.}
\label{fig:vrr}
\end{center}
\end{figure*}

Our main aim is to acquire a better insight into the structures which are formed
by our simulations. The two variables of interest for us are the density and radial velocity of the wind throughout the equatorial plane.
Following Fig.~6 of CO96, we show the radial dependence of density for various azimuthal sectors in Fig.~\ref{fig:rhor} and that of radial velocity
in Fig.~\ref{fig:vrr}.

We immediately notice a few important properties from these figures. %Model 1 (and model 1A in particular) shows a lot of instability. This comes from the fact
%that in that model, the spots' brightness contrast is very high, 
%leading to a strong local overloading of the wind. In the case of model 1A, this even leads to unexpected profiles,
%which are divided into two parts: the inner wind follows a structure similar to that of the other models, but the outer wind (past about $3 R_{*}$) shows a similar
%behaviour to the dark spot simulations conducted by \citet{1996ApJ...462..469C} (see especially their Fig.~8).
%Moreover, an
An obvious effect that can be seen is that the density and velocity structures are weaker for models in which the spots are larger. 
This is due to the fact that the spot size,
under our set of constraints, is
anticorrelated with the brightness contrast. Therefore, for a given photometric variation amplitude, larger spots will generate more subtle structures in both 
velocity and density.
Especially in the case of model 3, these structures produce barely noticeable density perturbations.

\begin{figure*}
\begin{center}
%\vspace{3.5cm}
\subfigure[Model 1A]{\includegraphics[width=3.4in]{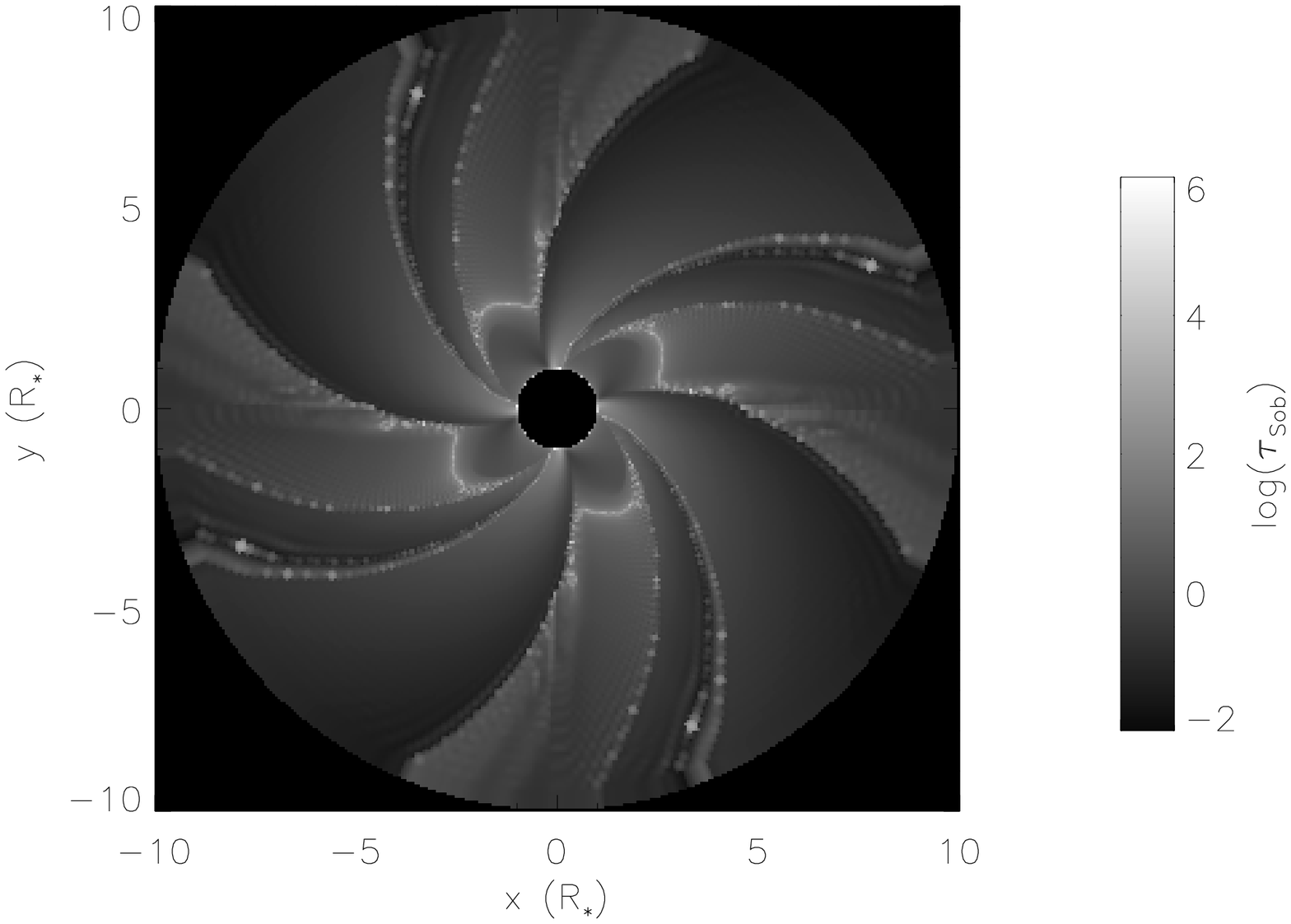}}
\subfigure[Model 1B]{\includegraphics[width=3.4in]{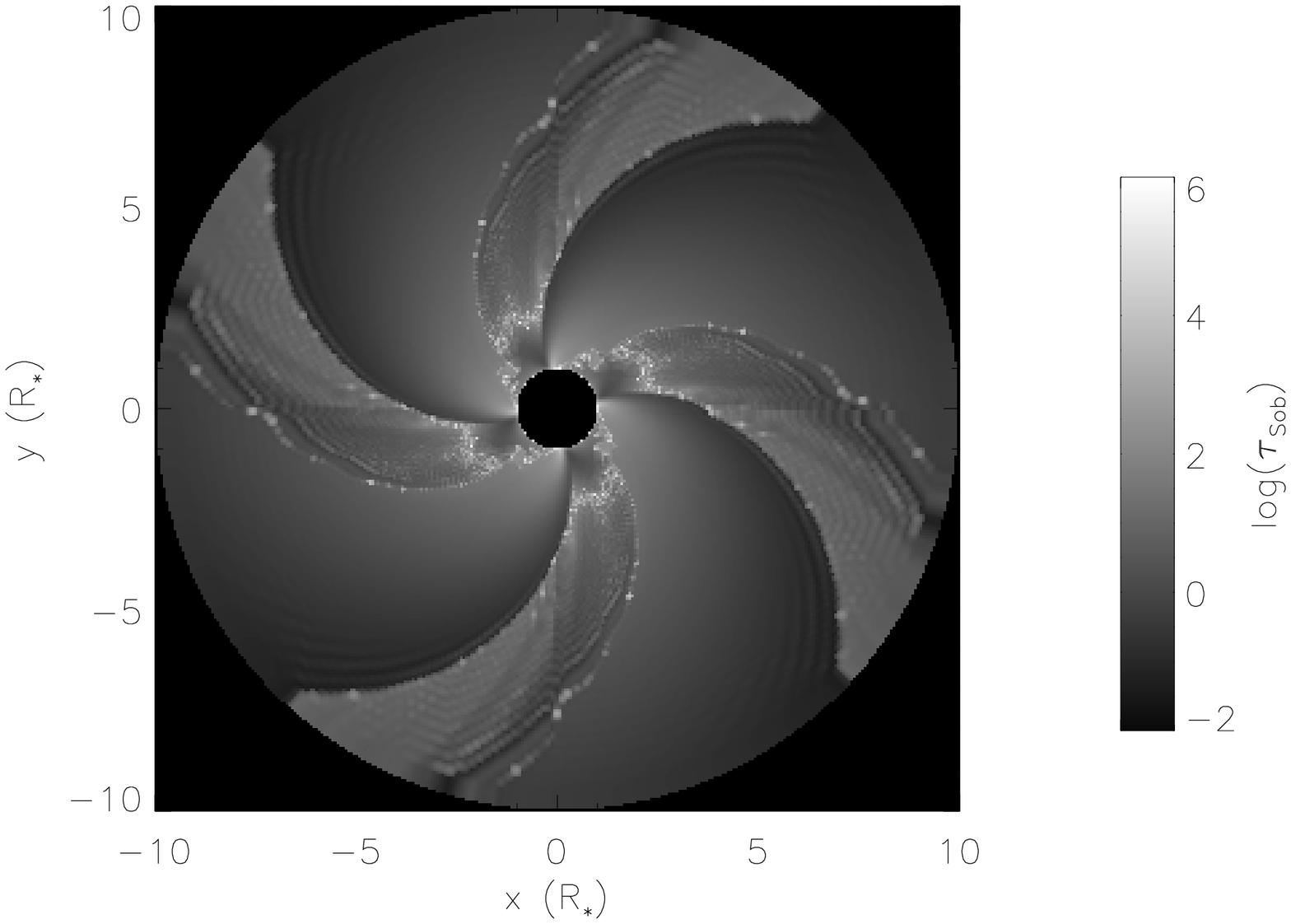}}
\subfigure[Model 2A]{\includegraphics[width=3.4in]{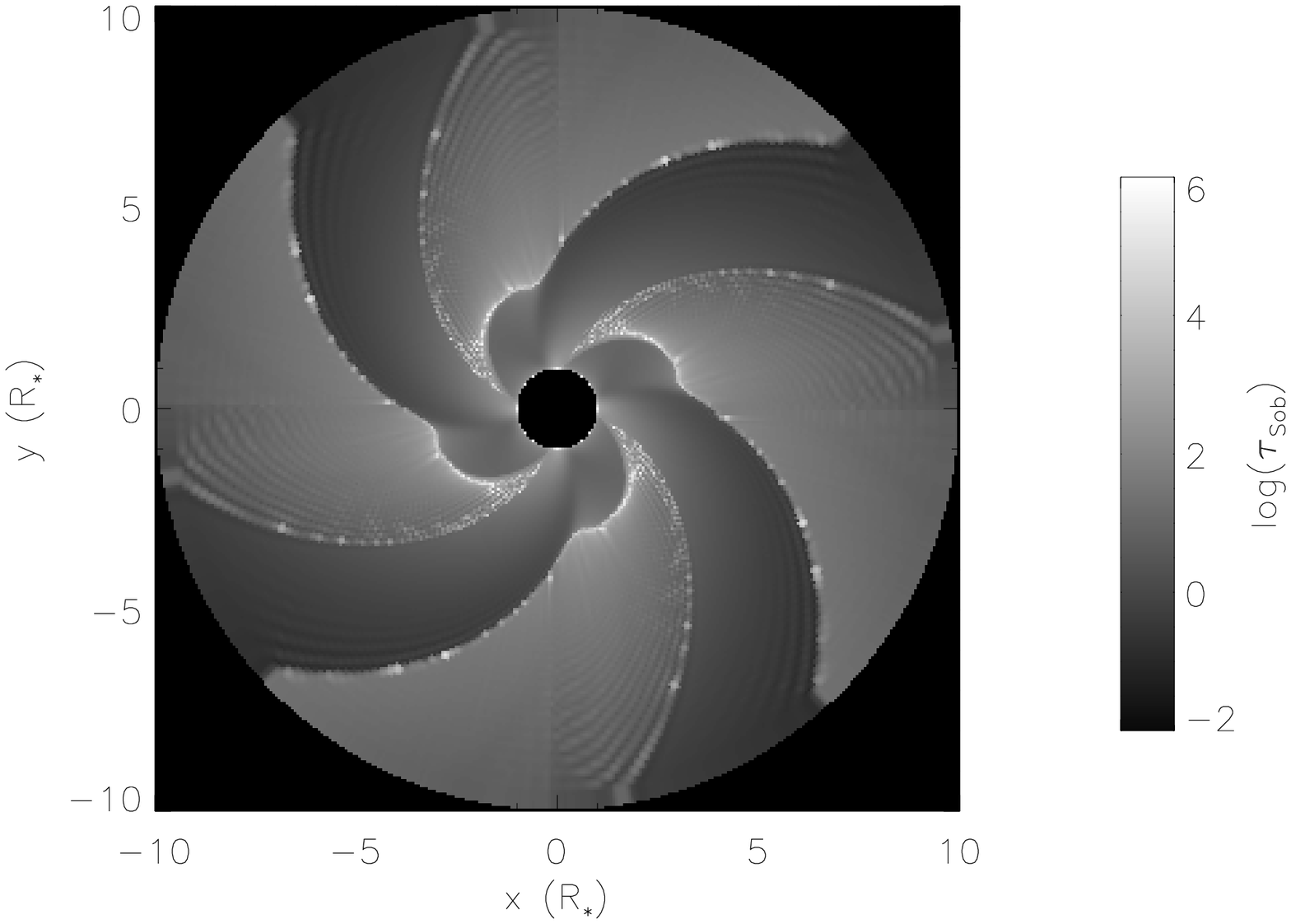}}
\subfigure[Model 2B]{\includegraphics[width=3.4in]{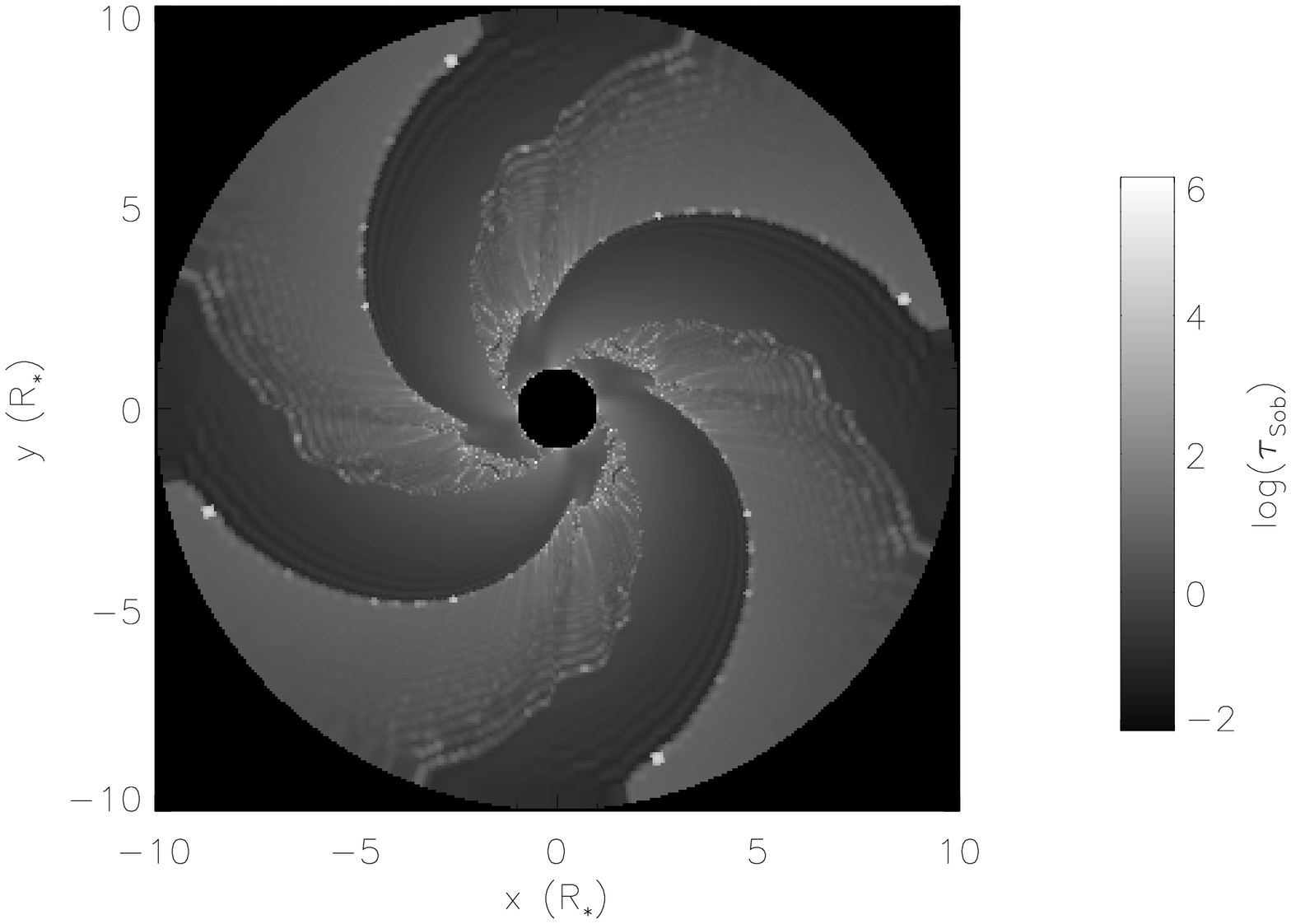}}
\subfigure[Model 3A]{\includegraphics[width=3.4in]{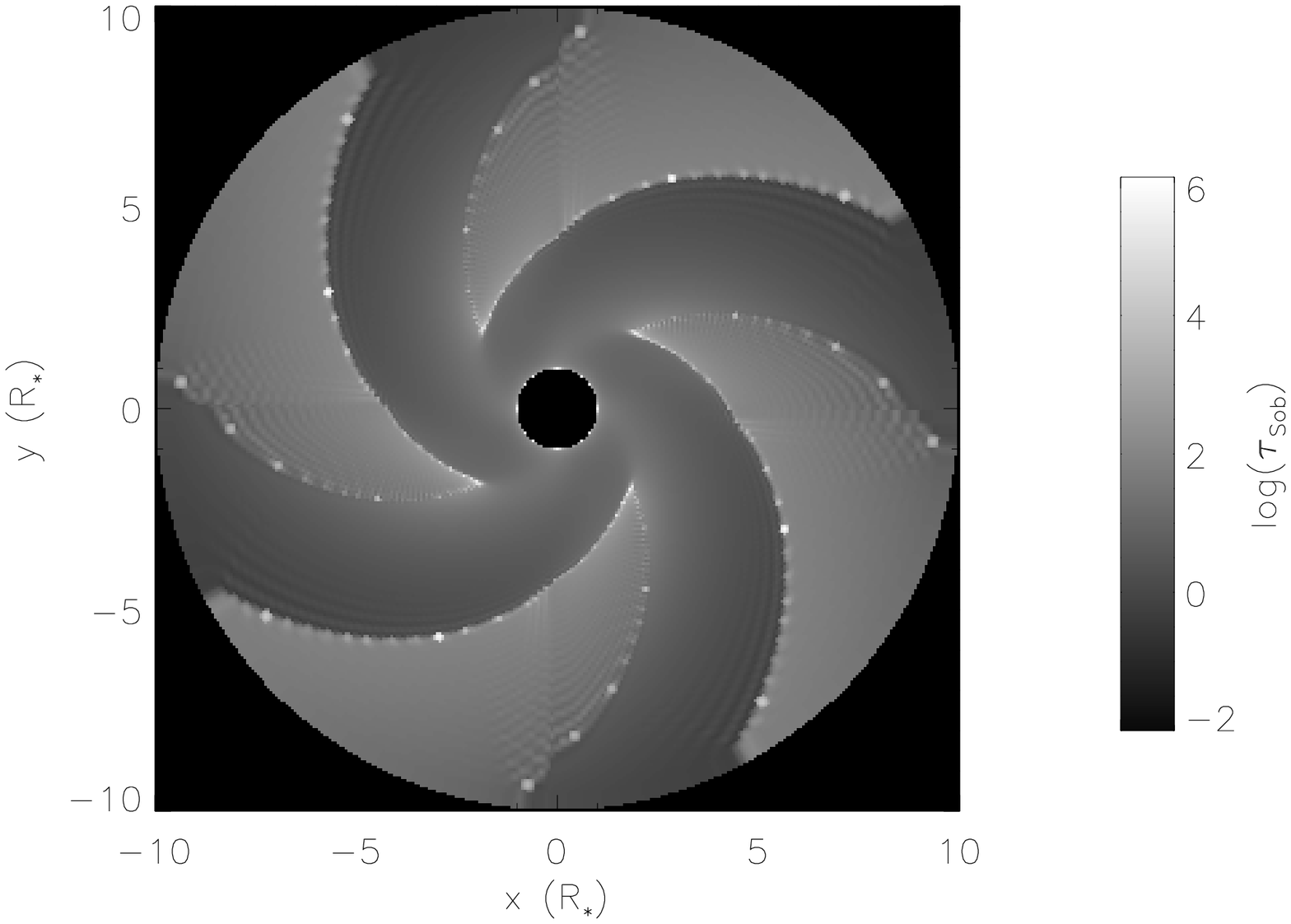}}
\subfigure[Model 3B]{\includegraphics[width=3.4in]{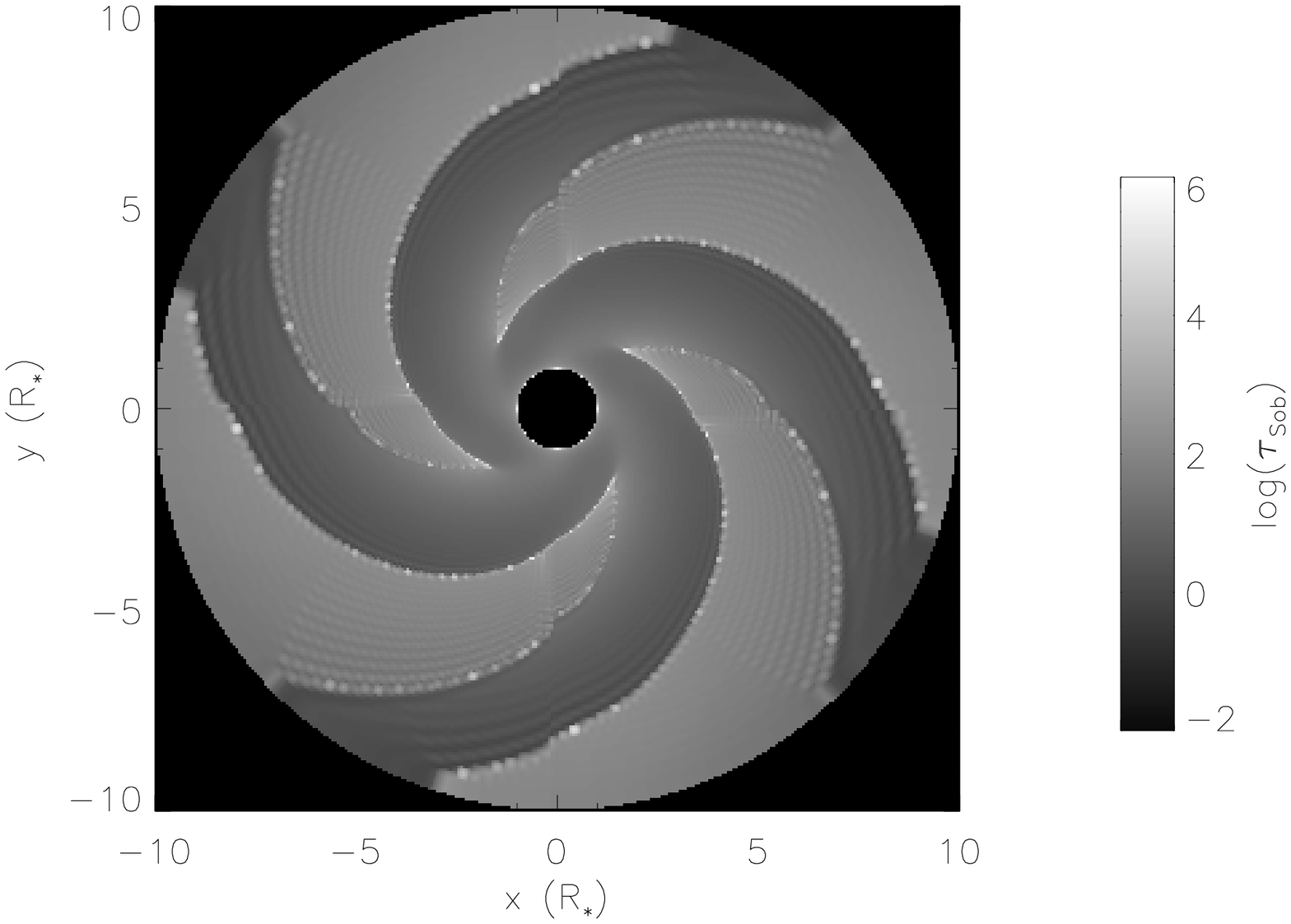}}
\caption{Equatorial plane greyscale visualizations of $\log(\tau_{\textrm{Sob}})$, the Sobolev (radial) optical depth, indicating where in the wind
the absorption occurs in each model.}
\label{fig:rhodrbdu}
\end{center}
\end{figure*}

Another useful visualization of these results is to compute wind optical depth, in order to identify 
the regions in the wind where most of the absorption takes place. 
In the Sobolev approximation \citep{1960mes..book.....S}, the radial optical depth can be written as:

\begin{equation}
\tau_{\textrm{Sob}} = \frac{\rho q \kappa_{\textrm{e}} c}{dv/dr}
\end{equation}

\noindent where $q$ is the frequency-integrated line strength, %\footnote{The line strength $q$ is related to $\kappa_{0}$ as follows:
%
%\begin{equation}
%q = \frac{\kappa_{0}}{\kappa_{\textrm{e}}}\frac{v_{\textrm{th}}}{c}
%\end{equation}
%
%\noindent where $v_{\textrm{th}}$ is the ion thermal velocity.},
$\kappa_{\textrm{e}}$ is the Thomson scattering opacity, $c$ is the speed of light
and $dv/dr$ is the radial velocity gradient. %In other words, the optical depth is proportional to the density divided by the radial velocity gradient.
We compute this quantity throughout the wind and plot it for each model in Fig.~\ref{fig:rhodrbdu}.

Models 1A and 1B lead to particularly interesting structures: the curvature of the inner edge of the CIRs is very pronounced in the inner wind. This suggests
that the extremely overluminous spots overload the wind so strongly that material would perhaps fall back onto the star if it were not for the ``boost"
provided by the adjacent spot as the star rotates. This behaviour might therefore be a consequence of the chosen number of spots and the
associated boundary conditions. However, an investigation of the influence of different spot distributions on this phenomenon is not within the scope
of this study.
%Once again, models 1A and 1B lead to somewhat unstable structures, whereas models with larger spots 
%generate fairly well-behaved absorption zones. This effect is very
%similar to that observed in the 1D density and radial velocity profiles.
We also notice that the larger, weaker spots produce absorption further from the stellar surface. Therefore, in terms of reproducing the observed properties
of DACs, there is a trade-off between having spots that are large enough to produce wind structures that cover a significant fraction of the stellar
disk near the surface, and brightness contrasts that are strong enough to overload the wind
in such a way that the velocity kinks (or breaks in the radial velocity profile when the acceleration becomes null or negative which are responsible
for the DACs, \citealt{1996ApJ...462..469C}) are close enough to the surface that their absorption is noticeable.

\subsection{Synthetic line profiles}\label{ssec:seiresults}

Finally, we compute synthetic resonance line profiles for each of our 6 models. To reduce calculation time, we only computed 14 different phases over a quarter
of a rotation period, and then since our simulations are in steady-state, plotted them twice 
to show two structures evolving in the dynamic spectra. Then, we approximate a ``least-absorption" template
spectrum, which corresponds roughly to the expected profile without the extra absorption caused by the CIRs \citep{1987ApJ...317..389P}, 
by using the maximum value of intensity among the 14 phases for each velocity bin. This constitutes our reference spectrum, by which we divide
each spectrum to obtain a quotient spectrum, which is plotted against phase to create our dynamic spectra, following the usual procedure
used to visualize DACs in observational data (e.g., \citealt{1996A&AS..116..257K}).

Once the dynamic quotient spectra are computed, we compare their characteristics with those of observed UV resonance line profiles
from $\xi$ Persei. In particular, we base our comparison on the quantitative analysis performed by \citet{1999A&A...344..231K}. Their figure 6 shows that the DACs' maximum
depth (or the minimum quotient flux) is about 20-30\%, and their figures 7 and 8 show that the DACs typically first appear in the profiles at about one-half 
the terminal velocity.% A variety of DAC morphologies are shown in the 4 timeseries which are analyzed in that study.

\begin{landscape}
\begin{figure}
\begin{center}
\subfigure[Model 1A synthetic line profile.]{\includegraphics[width=4.7in]{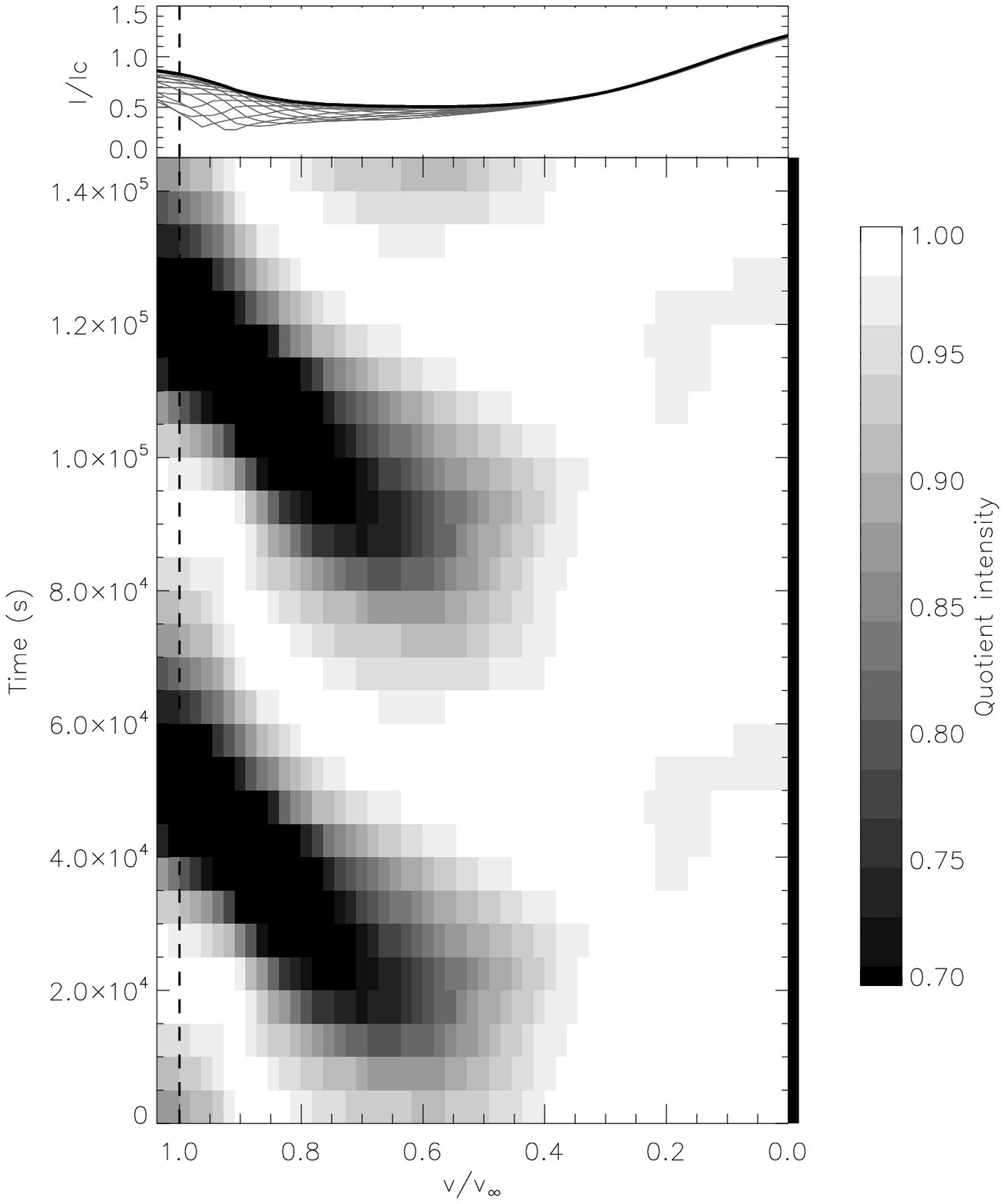}}
\subfigure[Model 1B synthetic line profile.]{\includegraphics[width=4.7in]{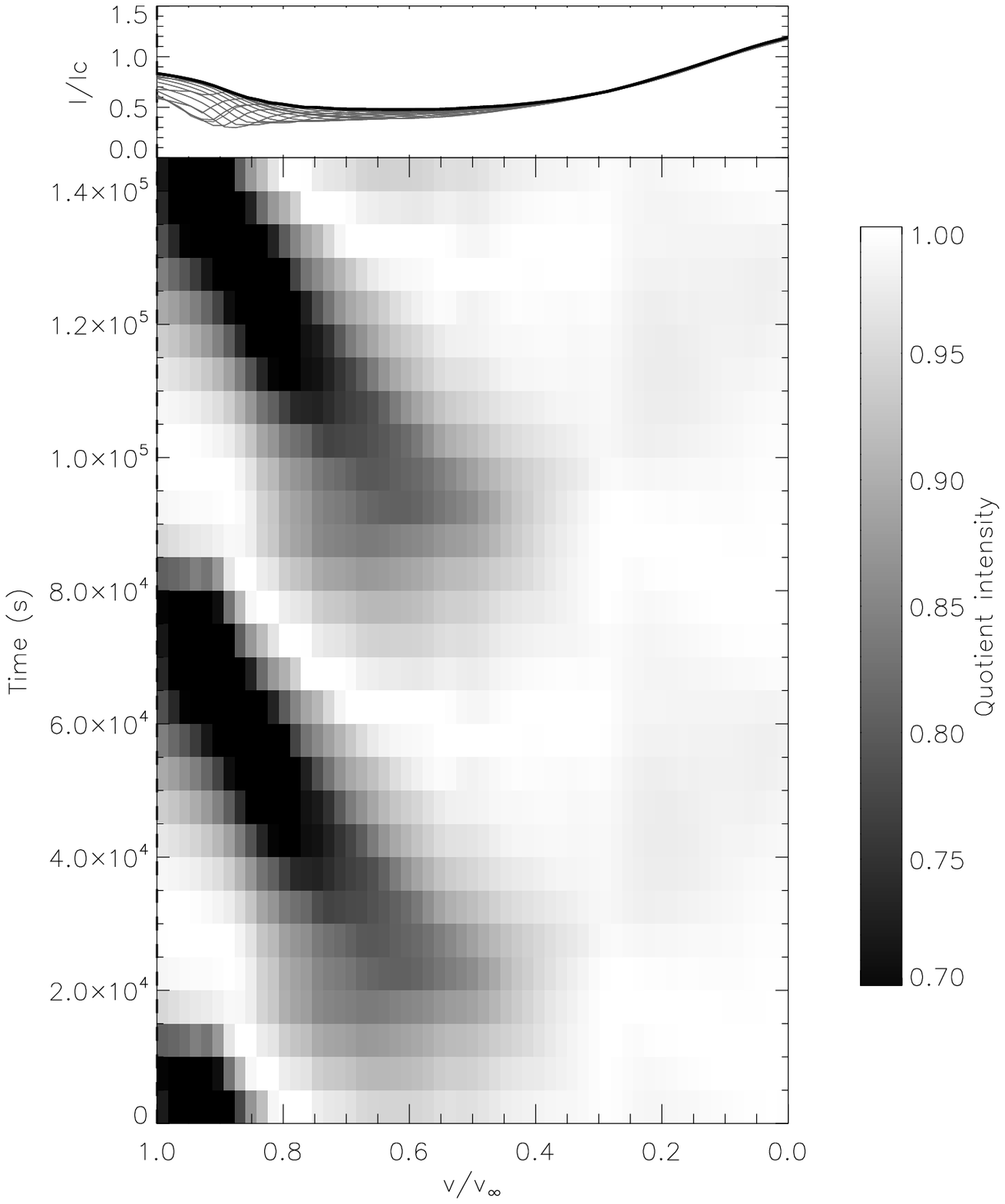}}
\caption{\textit{Left:} Dynamic spectrum (bottom panel) showing synthetic line profiles generated from model 1A (shown in the top panel in grey) divided by a
``least-absorption" template spectrum (shown in the top panel in black). The dashed black line represents $v_{\infty}$. \textit{Right:} Same, but for model 1B.
These dynamic spectra illustrate the fact that models 1A and 1B do not appear to appropriately reproduce the observed DAC behaviour.}
\label{fig:dynspec1}
\end{center}
\end{figure}
\end{landscape}

\begin{landscape}
\begin{figure}
\begin{center}
%\vspace{3.5cm}
\subfigure[Model 2A synthetic line profile.]{\includegraphics[width=4.7in]{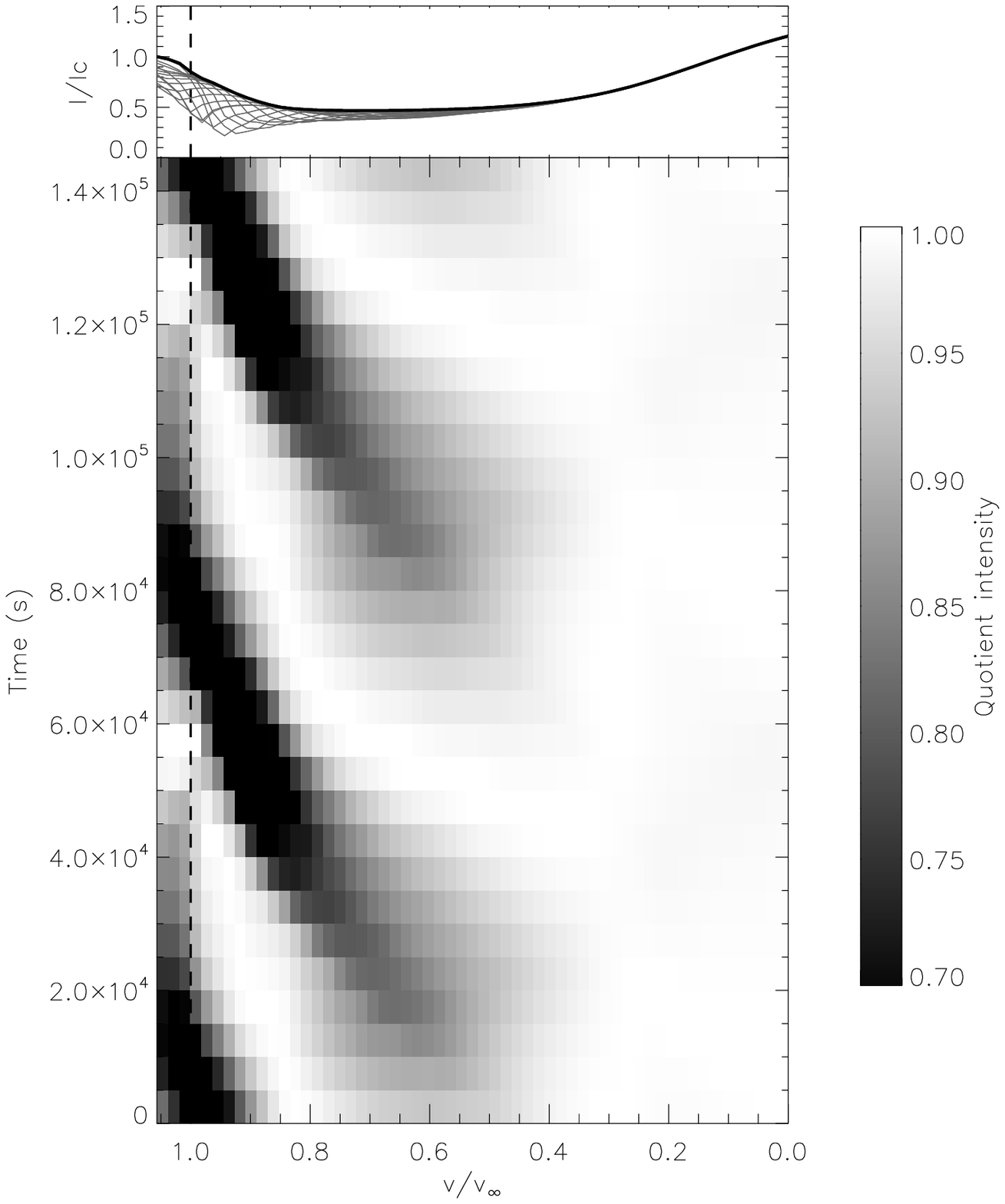}}
\subfigure[Model 2B synthetic line profile.]{\includegraphics[width=4.7in]{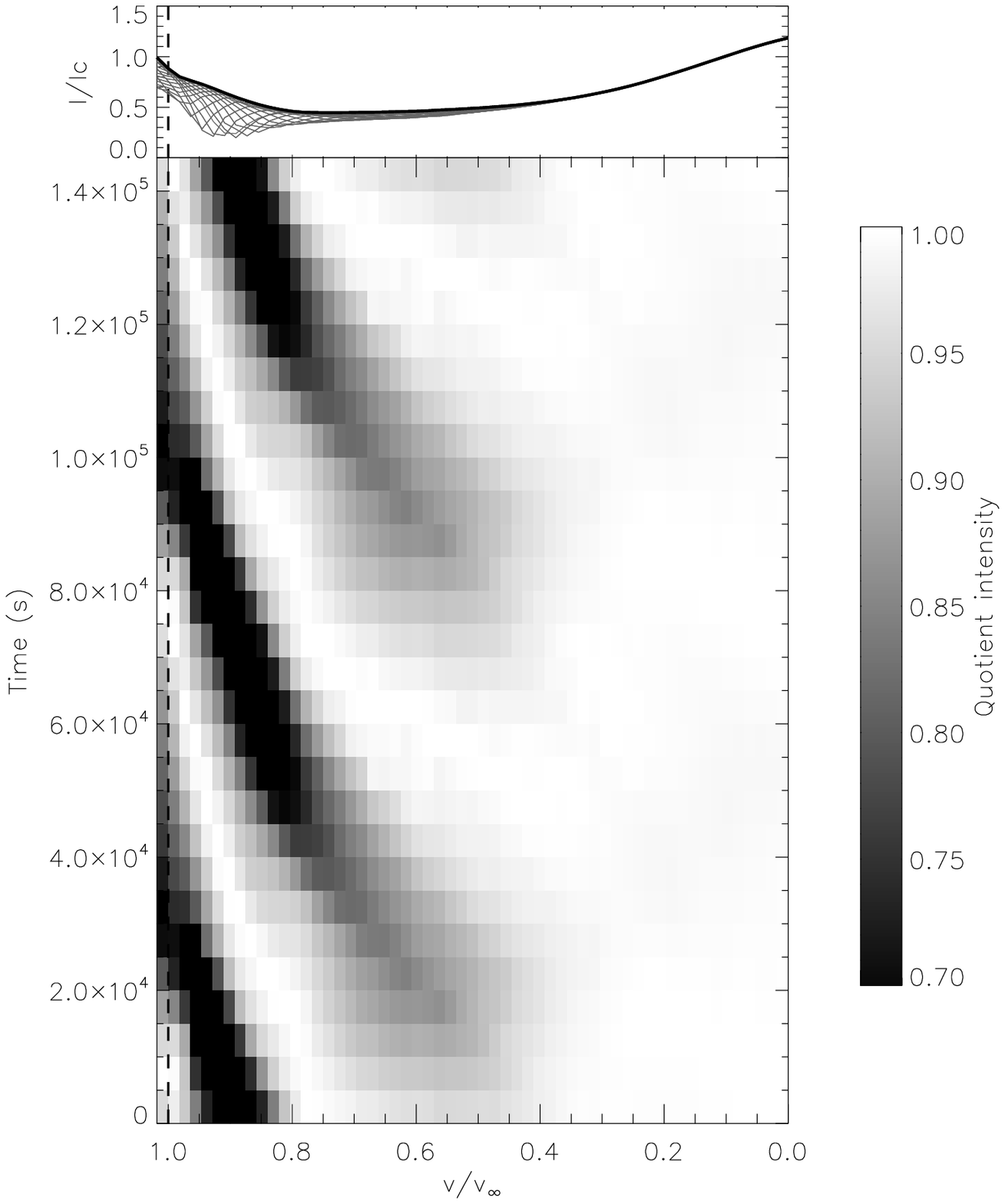}}
\caption{Same as Fig.~\ref{fig:dynspec1}, but for models 2A and 2B. In this case, the DACs behave in a way that is compatible with observations:
as they approach terminal velocity, their acceleration decreases and they become narrower.}
\label{fig:dynspec2}
\end{center}
\end{figure}
\end{landscape}

\begin{landscape}
\begin{figure}
\begin{center}
%\vspace{3.5cm}
\subfigure[Model 3A synthetic line profile.]{\includegraphics[width=4.7in]{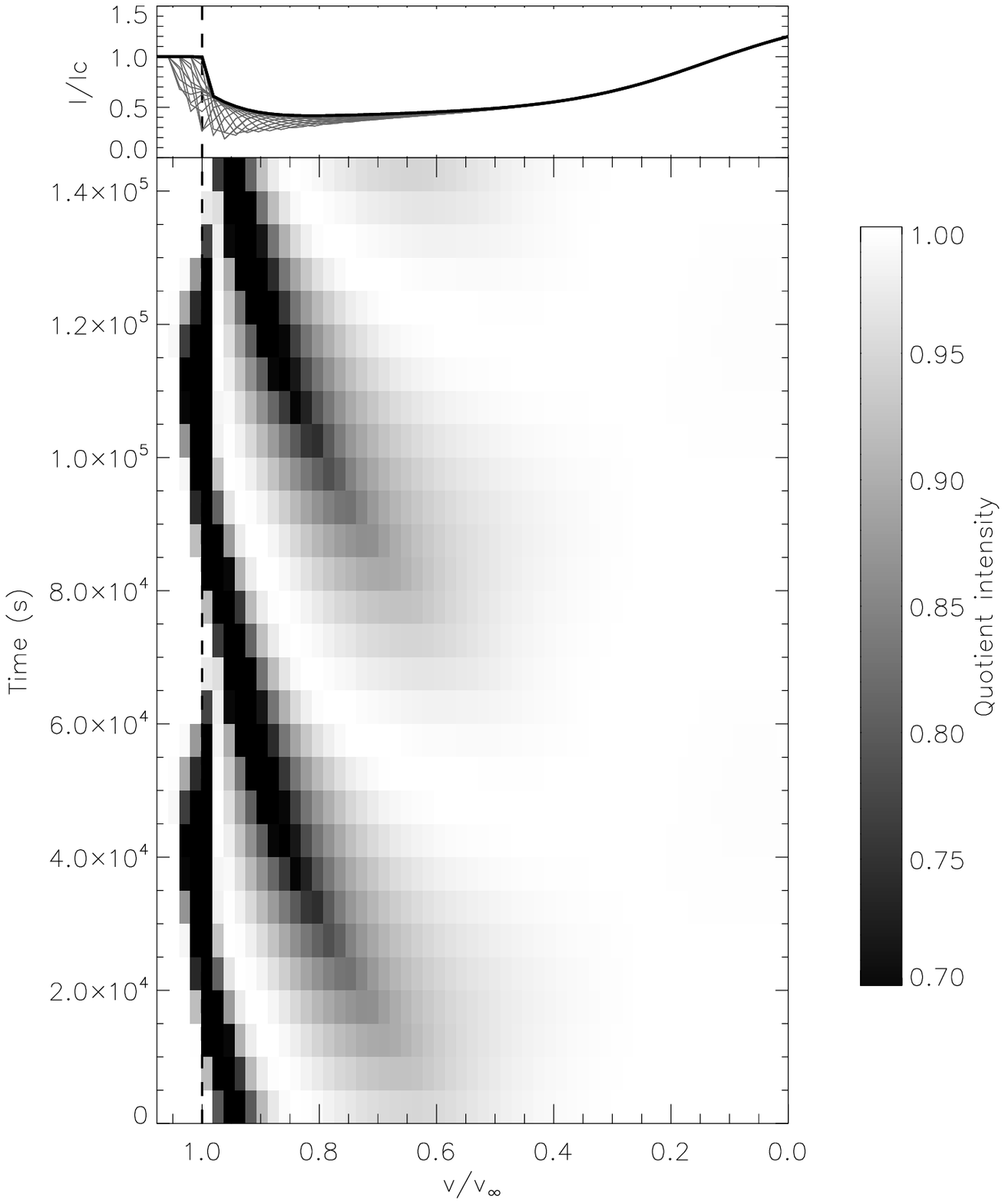}}
\subfigure[Model 3B synthetic line profile.]{\includegraphics[width=4.7in]{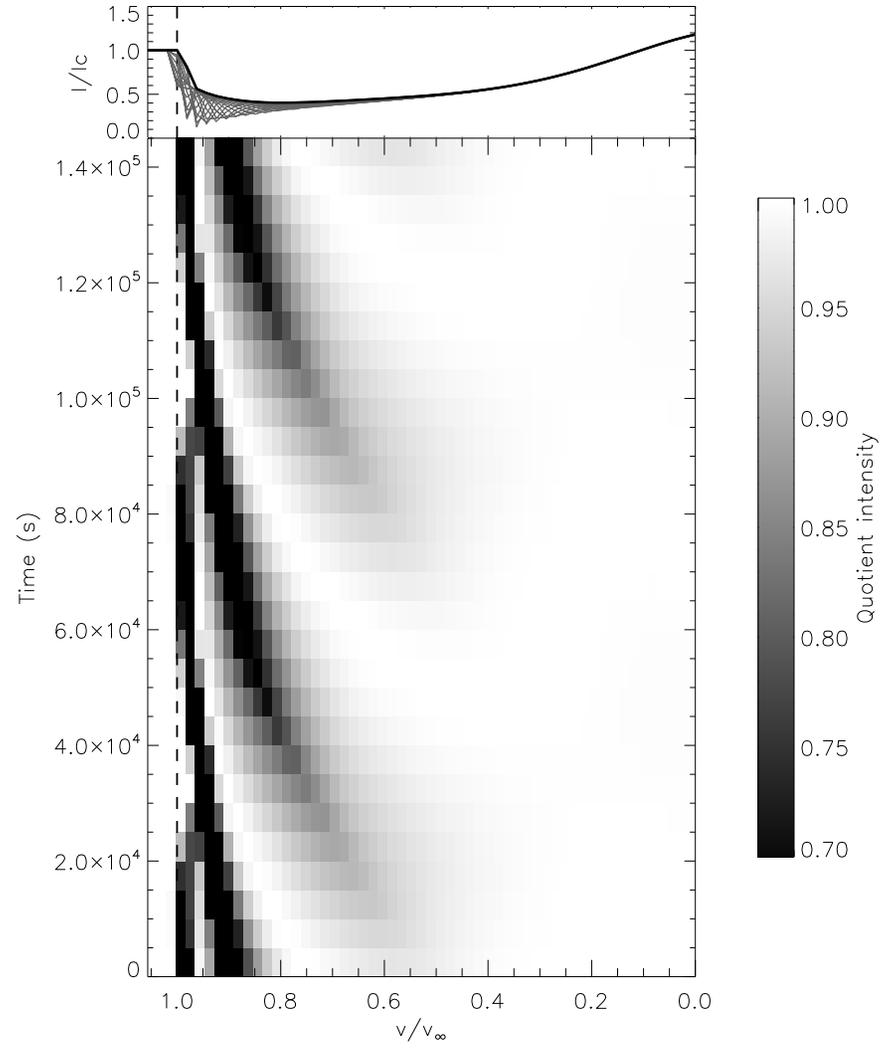}}
\caption{Same as Fig.~\ref{fig:dynspec1}, but for models 3A and 3B.}
\label{fig:dynspec3}
\end{center}
\end{figure}
\end{landscape}

The results of our SEI calculations are shown in Figs.~\ref{fig:dynspec1} to \ref{fig:dynspec3}. 
The terminal velocity that is shown on these figures and that is used as a scaling parameter to express the velocity range is estimated from the 
least-absorption template, in a manner aimed at best reproducing the observational procedure. 
Since this template was constructed using the maximum value of flux for each velocity bin, 
some structures extend slightly beyond the inferred terminal velocity. However, this does not particularly affect the scaling of the wind 
since the underestimation of the terminal velocities is small compared to the terminal velocities themselves. 

A first immediate conclusion derived from Figs.~\ref{fig:dynspec1} to~\ref{fig:dynspec3} is that not all models produce line profile
variations that are qualitatively compatible with those observed in the UV resonance lines of $\xi$ Persei. %Indeed,
%we produce signatures which are morphologically similar to those seen in observations. 
%In particular, model 1B produces a kind of
%feature consisting of two DACs accelerating at different rates and meeting at around terminal velocity (we will denote this feature as a ``forked DAC")
%that is occasionally seen in the UV resonance lines of $\xi$ Persei (most noticeably in the October 1994 run, at around $\textrm{BJD} = 2449644$ and 
%$\textrm{BJD} = 2449646$ in Fig.~4 of \citealt{2001A&A...368..601D}). Models 2A, 2B, 3A and 3B, on the other hand, produce rather classical DACs. 
Indeed, models 1A and 1B produce DACs that are morphologically quite different from those that are observed: unlike $\xi$ Persei's DACs, 
rather than getting narrower as they migrate towards terminal velocity, they remain very broad. Furthermore, the
observed DACs accelerate at a decreasing rate as they evolve through the velocity space to approach an ``asymptotic velocity"; in our case, models 
1A and 1B lead to DACs accelerating at an almost constant rate, extending into the blue edge of the line profile as the entire velocity structure
of the wind is perturbed. This is somewhat unsurprising, given the peculiar structures revealed in the equatorial plane optical depth plots of the winds of
both of these models (Fig. \ref{fig:rhodrbdu}). This finding might also be in line with the conclusion of \citet{2015ApJ...809...12M} that the spots responsible for
DACs must cover some significant portion of the stellar disk.

The remaining 4 models
seem to reproduce the DAC behaviour adequately.
This is quite remarkable,
especially for models 3A and 3B since, as mentioned earlier, the density and velocity structures 
found in the wind are rather subtle, which illustrates very well the conclusion
of \citet{1996ApJ...462..469C}: that DACs are formed by velocity kinks, not by overdense regions in the wind.

%\begin{landscape}

%\afterpage{
%\clearpage
%}

%\begin{figure*}
%\vbox to220mm{\vfil \includegraphics{dynspec1.ps}
%\caption{}
%\vfil}
%\label{landfig}
%\end{figure*}

%\begin{figure*}
%\begin{center}
%\vspace{3.5cm}
%\subfigure[Model 1A synthetic line profile.]{\includegraphics[width=3.0in]{new_sei_r05s132d00.eps}}
%\subfigure[Model 1B synthetic line profile.]{\includegraphics[width=3.0in]{new_sei_r05s132d10.eps}}
%\includegraphics[width=6.8in]{new_sei_r05s132d00.eps}
%\caption{Dynamic spectrum (bottom panel) generated using synthetic line profiles generated from model 1A (shown in the top panel in grey) divided by an
%``unperturbed" template spectrum (shown in the top panel in black). The dashed black line represents $v_{\infty}$.}
%\label{fig:dynspec1a}
%\end{center}
%\end{figure*}

%\begin{figure*}
%\begin{center}
%\vspace{3.5cm}
%\includegraphics[width=6.8in]{new_sei_r05s132d10.eps}
%\caption{Same as Fig.~\ref{fig:dynspec1a}, but for model 1B.}
%\label{fig:dynspec1b}
%\end{center}
%\end{figure*}

%\afterpage{
%\clearpage
%}

%\begin{figure*}
%\begin{center}
%\vspace{3.5cm}
%\includegraphics[width=6.8in]{new_sei_r10s033d10.eps}
%\caption{Same as Fig.~\ref{fig:dynspec1a}, but for model 2B.}
%\label{fig:dynspec2b}
%\end{center}
%\end{figure*}

%\afterpage{
%\clearpage
%}

%\begin{figure*}
%\begin{center}
%\vspace{3.5cm}
%\includegraphics[width=6.8in]{new_sei_r20s008d10.eps}
%\caption{Same as Fig.~\ref{fig:dynspec1a}, but for model 3B.}
%\label{fig:dynspec3b}
%\end{center}
%\end{figure*}

%\end{landscape}

As to the quantitative behaviour of these DACs, we see that for models 2A, 2B, 3A and 3B (Figs.~\ref{fig:dynspec2} and~\ref{fig:dynspec3}) 
DACs first appear at about one-half 
the terminal velocity. However, to better describe their behaviour, we can use the DAC-fitting method used by \citet{1983ApJ...268..807H} 
and \citet{1999A&A...344..231K} by fitting a Gaussian absorption profile to each absorption feature
in the quotient spectra:

\begin{equation}\label{eq:dacfit}
I(v) = \exp \left( -\tau_{c} \exp \left[ -\left( \frac{v-v_{c}}{v_{t}} \right)^2 \right] \right)\,,
\end{equation}

\noindent where $v_{c}$ corresponds to the central velocity of the feature, $v_{t}$ corresponds to its width and $\tau_{c}$ is the central optical depth.
This has the additional advantage that we infer the behaviour of DACs in our theoretical spectra using the same method of measurement as used for the observations.
We only apply this method to the four models which successfully reproduce DAC behaviour: 
indeed, not only is the DAC morphology at the blue edge of the line profile
problematic for models 1A and 1B, but we found that their DACs are much more non-Gaussian than those of other models throughout the velocity space, and as such,
much more difficult to model using this approach. Accordingly, we decided to forgo this analysis for models 1A and 1B.

\begin{figure}
\begin{center}
%\vspace{3.5cm}
\includegraphics[width=3.3in]{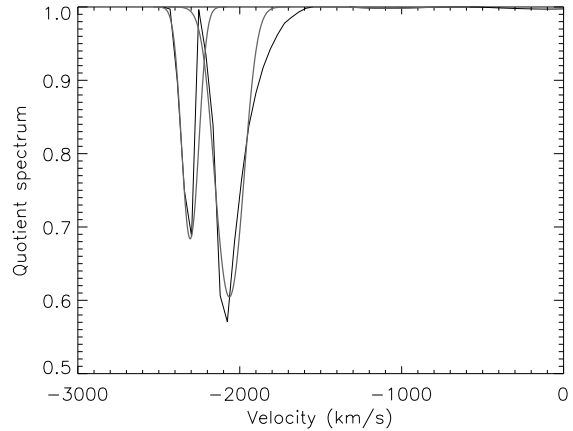}
\caption{Example of the ``appearance" of a DAC (seen here at about $v = 1100$ km/s) in model 3A. 
This corresponds to the first spectrum in which this DAC is discrete and isolated.
%from the lower-velocity absorption ($v < 600$ km/s). 
Two other, ``older" DACs can be seen to still exist in the profile. 
The fits to the DACs correspond to the thick gray lines; note that the higher velocity DACs are not perfectly fit since
they are not fully described by Gaussian curves.}
\label{fig:dac_example}
\end{center}
\end{figure}

We find the absorption feature for each model that appears at the lowest velocity across all phases and consider that velocity to be the starting velocity of the DACs
(expressed as a fraction of the terminal velocity of the reference spectrum in Table~\ref{tab:wprop}). %This procedure was complicated by the fact that in some
%spectra, there seems to be a broad, ill-defined low-velocity (encompassing $v = 0$) 
%absorption feature present. However, the criterion we used to define the ``appearance" of a DAC was to
%consider the spectrum in which this feature 
More specifically, a DAC is considered to ``appear" for the first time when it becomes discrete, i.e., when it can be fully isolated (or in other
words when the profile returns to the continuum on either side of the absorption feature). The starting velocities might be slightly underestimated
due to the chosen temporal resolution, but, in the context of this fitting method, the very small values of central optical depth measured when they first
appear suggests that this effect is very modest as they cannot have appeared much earlier or they would not have been observable\footnote{That is not to say that
the observed low-velocity absorption (e.g., in $\xi$ Persei, in the context of phase bowing; \citealt{1995ApJ...453L..37O,1997A&A...327..699F}) is unimportant, 
but it does not correspond to the DACs we are fitting and cannot be accurately described by a Gaussian profile, therefore it is not considered in this analysis.}. 
Therefore, we consider our current time sampling to be adequate.
An example of a spectrum in which this first happens (and the fitted DACs) is shown
in Fig.~\ref{fig:dac_example}.

\begin{figure}
\begin{center}
%\vspace{3.5cm}
\subfigure[DAC central velocity ($v_{c}$)]{\includegraphics[width=3.3in]{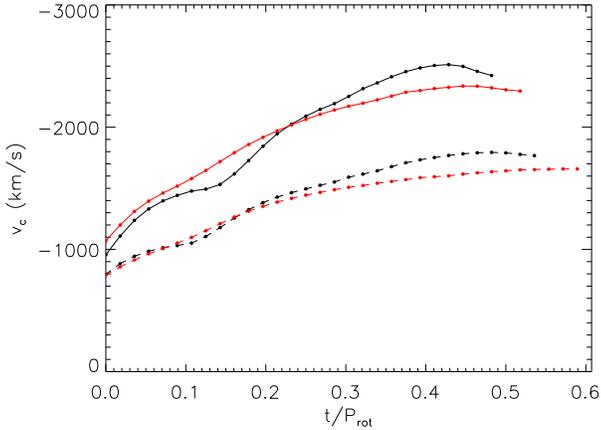}}
\subfigure[DAC central optical depth ($\tau_{c}$)]{\includegraphics[width=3.3in]{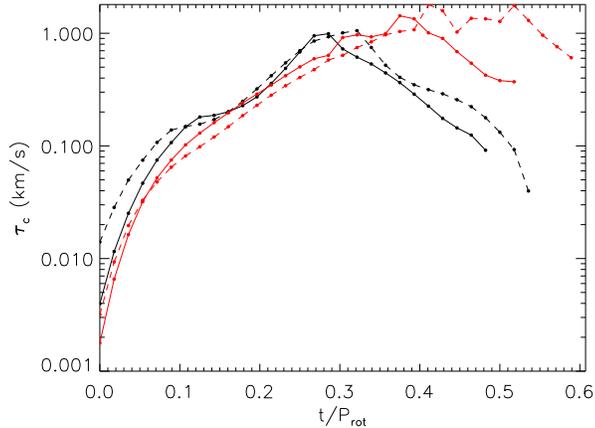}}
\subfigure[DAC width ($v_{t}$)]{\includegraphics[width=3.3in]{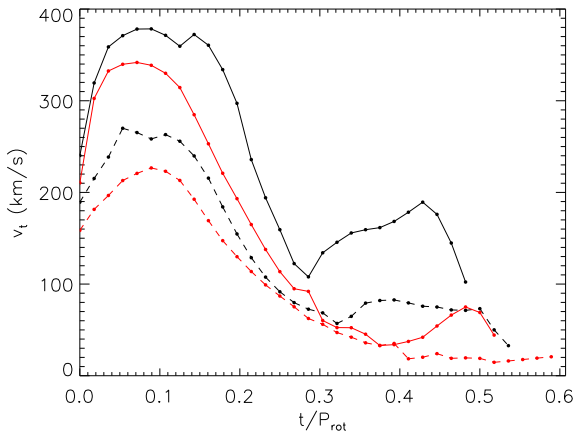}}
\caption{Characteristics of the DACs as a function of time; the black curves correspond to model 2 and the red curves to model 3.
Full curves correspond to the ``A" variety of each model (not including ionization effects) and the dashed curves correspond to the ``B" variety (including 
ionization effects). For all three panels, the horizontal axis corresponds to the time elapsed since the ``appearance" of the DAC, as a fraction of the rotation
period.}
\label{fig:dacs}
\end{center}
\end{figure}

The starting velocities reported in Table~\ref{tab:wprop} are compatible with 
those found for $\xi$ Persei by
\citet{1999A&A...344..231K}. Following their analysis, we can also trace the evolution of the 3 quantities appearing in Eq.~\ref{eq:dacfit} for a single DAC
caused by a CIR wrapping around
the star. Fig.~\ref{fig:dacs} shows the results of these calculations, respectively for $v_{c}$, $\tau_{c}$ and $v_{t}$.

%\begin{figure}
%\begin{center}
%\vspace{3.5cm}
%\includegraphics[width=3.3in]{dac.eps}
%\caption{DAC characteristics for model 2B: the top panel shows the evolution of the central velocity of the DAC ($v_{c}$) with time, 
%the middle panel traces the central optical depth of the absorption feature ($\tau_{c}$) and the bottom panel traces its broadening parameter ($v_{t}$).}
%\label{fig:dacs}
%\end{center}
%\end{figure}

%\begin{landscape}
%\end{landscape}

%\begin{figure}
%\begin{center}
%\vspace{3.5cm}
%\includegraphics[width=3.3in]{dac_velc.eps}
%\caption{Central velocity ($v_{c}$) of the DACs as a function of time; the black curves correspond to model 1, red curves to model 2 and blue curves to model 3.
%Full curves correspond to the ``A" variety of each model (not including ionization effects) and the dashed curves correspond to the ``B" variety (including 
%ionization effects).}
%\label{fig:dac_velc}
%\end{center}
%\end{figure}

%\begin{figure}
%\begin{center}
%\vspace{3.5cm}
%\includegraphics[width=3.3in]{dac_tauc.eps}
%\caption{Same as Figure~\ref{fig:dac_velc}, but tracing instead the central optical depth of the DACs ($\tau_{c}$).}
%\label{fig:dac_tauc}
%\end{center}
%\end{figure}

%\begin{figure}
%\begin{center}
%\vspace{3.5cm}
%\includegraphics[width=3.3in]{dac_velt.eps}
%\caption{Same as Figure~\ref{fig:dac_velc}, but tracing instead the width of the DACs ($v_{t}$).}
%\label{fig:dac_velt}
%\end{center}
%\end{figure}

The fit parameters of our synthetic DACs show clear trends: 
the DACs become stronger and narrower as they evolve through the velocity space. These trends are less clear
at low and near-terminal velocity. When the DACs first form, there is significant asymmetry in their profile, 
leading to unsatisfactory fits using a purely Gaussian profile; furthermore, it is more difficult to fit weak features. 
When they are near terminal velocity, the departures from the trends are due in part to the finite simulation range, which truncates the
CIRs which cause these features. ``Quasi-blending" also becomes an issue towards terminal velocity, as two DACs nearly overlap in velocity range, causing 
slight problems with the fit\footnote{Of course, given the radial velocity structure of the wind, two DACs can never actually blend, but
the Gaussians used to fit them can overlap.}.

Finally, the maximum depth of the quotient dynamic spectrum for each model is compiled in Table~\ref{tab:wprop}. Once again, these values are compatible with the
observed values, which range between about 20 and 30\%. There seems to be a slight trend between maximum depth and spot size; larger spots lead to
somewhat deeper DACs. This might bear interesting consequences for a few other stars in the \citet{1996A&AS..116..257K} sample, which had nearly
saturated DACs. In order to achieve such deep DACs, the wind structures causing the absorption probably have to originate from very large surface spots
(in accordance with \citealt{2015ApJ...809...12M}), unless the stars in question exhibit a larger photometric variability than $\xi$ Persei.

%However, it should be noted that given the way our reference spectrum was constructed, the depths might be slightly overestimated,
%compared to the observational ones. Nevertheless, they span the expected range, confirming that we approximately reproduce the 
%quantitative properties of $\xi$ Persei's DACs.
In conclusion, by fitting our DACs using the same method used for the observed spectra of $\xi$ Persei, we conclude that the DACs produced by four of our 
models (2A, 2B, 3A and 3B) have quantitative properties that are very similar to those of $\xi$ Persei's DACs.

\subsection{Effect of ionization}

The most obvious effect observed when the ionization parameter $\delta$ is introduced is a global reduction of the wind velocity 
(as illustrated in Fig.~\ref{fig:vrr}), as well as an increase in the mass-loss rate (see Table~\ref{tab:wprop}).
We also see from Fig.~\ref{fig:rhodrbdu} that including a non-zero ionization parameter increases the optical depth of the CIRs 
and results in their appearance closer to the stellar surface.
However, the inclusion of this effect barely influences the computed line profiles. The main reason for this lies in the fact that for a
strong UV resonance line, the absorption is nearly saturated at low velocities, making it difficult to distinguish any additional absorption. Indeed, looking at
a typical unperturbed profile, we see that the absorption is already nearly saturated at about $v/v_{\infty} = 0.3$ 
(as shown in Fig.~\ref{fig:abs}). We do notice that, in Table~\ref{tab:wprop}, the `B' models typically have slightly deeper absorption than the corresponding
`A' models, but that effect is very small. %This means that resonance lines
%are not optimally suited to study the wind structure near the stellar surface, which is already well known (e.g., \citealt{2015ApJ...809...12M}). To fully account for that
%region of the wind, we would need to study synthetic excited state lines\footnote{It is also possible to investigate weaker lines, but this
%study specifically focussed on the type of lines in which DACs were observed.}, which our code cannot generate in its present version. 
The DACs produced in the `B' models are also systematically narrower than their counterparts computed from the `A' models, but once again this is not
very surprising since the entire velocity structure of the wind is scaled to a lower terminal velocity.
Modelling weaker resonance lines, as well as excited state lines (such as those investigated by \citealt{2015ApJ...809...12M}), 
might help in the future to define the role of the ionization parameter in our simulations.
%; however,
%further testing is required and considered to be outside the scope of this study. Therefore, 
%based solely on the direct comparison of our models with the observables at our disposal, little can be said at this time about the influence of
%the ionization parameter in our simulations.

\section{Conclusions and future work}\label{sec:concl}

In this study, we have carried out 2D radiation hydrodynamical simulations of $\xi$ Persei's wind and aimed to explore whether bright spots compatible with
recent observation could be responsible for its observed DACs. The spot
characteristics were constrained by the photometric limits found by \citet{2014MNRAS.441..910R}. 
We then extrapolated these results to 3D using the same prescription as CO96 and used 
SEI to generate UV resonance line profiles. This procedure allowed us to test whether various spot and wind parameters produced qualitatively similar 
wind structures, and whether the associated spectroscopic signatures were consistent with the DAC phenomenology. We used three sets of spot parameters (size and
amplitude), each of which was then tested with or without including the effects of the radial dependence of ionization levels in the wind.
Based on these experiments, we conclude the following:

\begin{itemize}

\item All 6 spot models cause perturbations in the velocity and density profiles of the wind. Models with smaller and stronger spots cause
structures with larger overdensities and velocity plateaus in the wind, 
while larger and weaker spots generate subtler structures. The hydrodynamical simulations also show evidence that the spots enhance the
overall mass-loss rate and that, as expected, 
the inclusion of the effect of radially-varying ionization globally slows down the wind and increases mass loss, leading to stronger
absorption in the CIRs.

\item The synthetic line profiles show classical DAC behaviour for 4 out of the 6 models.
%The synthetic line profiles show a variety of patterns of variability. In particular, the profile generated using model 1A does not reproduce DACs, 
%and models 2A, 2B, 3A and 3B successfully lead to classical DACs. Model 1B is an interesting case: the inferred synthetic line profiles show evidence
%of ``forked DACs", which are seen in a few timeseries of observations of $\xi$ Persei. For all 5 successful models, 
The quantitative behaviour of the DACs
also agrees with observations: they appear at a little less than one-half the terminal velocity, then become deeper and narrower as they evolve in the velocity
space, reaching a maximum depth of around 30\% for model 2A, going down to about 20\% for model 3B. 
Thus, the key result of this study is that we have successfully linked
the behaviours of two sets of
observations, i.e., optical photometry and UV spectroscopy, within the ``bright spot paradigm" introduced by CO96 by using $\xi$ Persei as a testbed.

\item The fact that models corresponding to a range of spot sizes reproduce the pattern of variability 
associated with DACs suggests that there can exist a variety of these structures
%The fact that different models reproduce different patterns of variability which are all seen in the observational data suggests that a variety
%of structures exist 
in the wind of $\xi$ Persei at different times, since there is no reason to favour one model over the other (and the quality of the existing data might not
allow us at this point to distinguish between these). However, our results also eliminate some portions
of the parameter space, thus allowing us to constrain this problem.
If we accept the idea that these structures are generated by bright spots
on the surface of the star, we deem our findings to be compatible with the idea that these spots are stochastic, appearing and
disappearing over time, with varying strengths and sizes. Such a scenario might explain the cyclical 
(rather than periodic) nature of DACs.

\end{itemize}

The main limitation of this study is that we have not convincingly addressed the role of the ionization factor in producing DACs. While its effect on the
global properties of the wind was quite obvious (and unsurprising), its inclusion did not seem to be a significant factor when it comes to the DACs' properties.
Further investigation will be required in order to clarify this issue.

As to the nature of the physical phenomenon giving rise to these spots, the present work helps place lower limits on the strength that magnetic spots 
(e.g., \citealt{2011A&A...534A.140C})
would need to have to produce the appropriate brightness enhancements. Using Eq.~5 from \citet{2014MNRAS.444..429D} and assuming $\log(g) = 3.5$, 
we approximate that the field strength required to generate
the brightness contrast used in model 2 is 360 G, whereas
model 3 requires a 160 G field. Assuming a best case scenario involving a spot situated in the center of the disk, and taking 
the field to be along the line of sight at each point (which is a valid approximation given the size of the spots), we expect to measure disk-averaged longitudinal
fields of, respectively, 11 G and 19 G. Such fields could be detectable using deep magnetometry; currently, the best longitudinal field error bar
obtained for $\xi$ Persei using NARVAL observations is 21 G \citep{2014MNRAS.444..429D}. 
For a given photometric variation constraint, larger spots would therefore yield larger longitudinal field values. However, as seen
throughout Section~\ref{sec:results}, larger spots lead to weaker CIRs, and the extra absorption they cause occurs further away from the star, which
might be at odds with various other observational diagnostics, including excited state lines \citep{2015ApJ...809...12M}.%; this means that the more likely 
%progenitors of the structures which cause DACs are also more likely to be detectable.

The best way to compute the behaviour of wind structures near the stellar surface would therefore be to generate synthetic line profiles that are
more sensitive to that region of the wind.
Another important observable to investigate would also be the behaviour of H$\alpha$. Indeed, variations in H$\alpha$ are known to be linked to DACs,
but their patterns typically do not look as well organized as that of their UV counterparts. Nevertheless, since high-resolution optical spectroscopy is
much more accessible than UV spectroscopy these days, it would be very useful to make sense of these patterns and determine whether they allow us to infer
anything about the surface perturbations which cause them.

Another extension of this study would be to investigate if our modeled CIRs can account for the X-ray variability observed in a number of O
stars, such as $\lambda$ Cephei \citep{2015A&A...580A..59R}. Currently our simulations are isothermal: we would need to track the temperature structure
of the wind and ideally produce 3D models to fully account for the modulation of the X-ray absorption by large-scale structures such as CIRs.

Finally, while this paper self-consistently accounts for the behaviour of two observables within a ``CO96-like" paradigm for a specific star, 
future work should extend the parameter space to account for the great
variety of DAC signatures found in all OB stars, not just $\xi$ Persei. 

\section*{Acknowledgments}

This research has made use of the SIMBAD database operated at CDS, Strasbourg, France and 
NASA's Astrophysics Data System (ADS) Bibliographic Services.

ADU gratefully acknowledges the support of the \textit{Fonds qu\'{e}b\'{e}cois de la recherche sur la nature et les technologies}. GAW is supported by an NSERC
Discovery Grant. JOS acknowledges funding from the European Union's Horizon 2020 research and innovation programme under the Marie Sklodowska-Curie grant agreement
no. 656725.

ADU also warmly thanks Alex Fullerton and V\'{e}ronique Petit for their helpful comments and kind guidance. We also wish to thank the anonymous reviewer who provided
very useful observations and suggestions which contributed to improve this paper and its impact.

\label{lastpage}


\begin{thebibliography}{99}
\bibitem[\protect\citeauthoryear{Abbott}{1982}]{1982ApJ...259..282A} Abbott D.C., 1982, ApJ, 259, 282
\bibitem[\protect\citeauthoryear{Cantiello \& Braithwaite}{2011}]{2011A&A...534A.140C} Cantiello M., Braithwaite J., 2011, A\&A, 534, A140
\bibitem[\protect\citeauthoryear{Castor, Abbott \& Klein}{1975}]{1975ApJ...195..157C} Castor J.I., Abbott D.C., Klein R.I., 1975, ApJ, 195, 157
\bibitem[\protect\citeauthoryear{Chen\'{e} et al.}{2011}]{2011ApJ...735...34C} Chen\'{e} A.-N., Moffat A.F.J., 
Cameron C., Fahed R., Gamen R.C., Lef\`{e}vre L., Rowe J.F.,
St-Louis N., Muntean V., de la Chevroti\`{e}re A., Guenther D.B., Kuschnig R., Matthews J.M., Rucinski S.M., Sasselov D., Weiss W.W., 2011, ApJ, 735, 34
\bibitem[\protect\citeauthoryear{Colella \& Woodward}{1984}]{1984JCoPh..54..174C} Colella P., Woodward P.R., 1984, \textit{J. Comp. Phys.}, 54, 174
\bibitem[\protect\citeauthoryear{Cranmer \& Owocki}{1995}]{1995ApJ...440..308C} Cranmer S.R., Owocki S.P., 1995, ApJ, 440, 308
\bibitem[\protect\citeauthoryear{Cranmer \& Owocki}{1996}]{1996ApJ...462..469C} Cranmer S.R., Owocki S.P., 1996, ApJ, 462, 469
\bibitem[\protect\citeauthoryear{David-Uraz et al.}{2012}]{2012MNRAS.426.1720D} David-Uraz A., Moffat A.F.J., Chen\'{e} A.-N., Rowe J.F., Lange N., Guenther D.B.,
Kuschnig R., Matthews J.M., Rucinski S.M., Sasselov D., Weiss W.W., 2012, MNRAS, 426, 1720
\bibitem[\protect\citeauthoryear{David-Uraz et al.}{2014}]{2014MNRAS.444..429D} David-Uraz A., 
Wade G.A., Petit V., ud-Doula A., Sundqvist J.O., Grunhut J., Shultz M.,
Neiner C., Alecian E., Henrichs H.F., Bouret J.-C., MiMeS Collaboration, 2014, MNRAS, 444, 429
\bibitem[\protect\citeauthoryear{de Jong et al.}{1999}]{1999A&A...345..172D} de Jong J.A., Henrichs H.F., Schrijvers C., Gies D.R., Telting J.H.,
Kaper L., Zwarthoed G.A.A., 1999, A\&A, 345, 172
\bibitem[\protect\citeauthoryear{de Jong et al.}{2001}]{2001A&A...368..601D} de Jong J.A., Henrichs H.F., Kaper L., Nichols J.S., Bjorkman K., Bohlender D.A., 
Cao H., Gordon K., Hill G., Jiang Y., Kolka I., Morrison N., Neff J., O'Neal D., Scheers B., Telting J.H., 2001, A\&A, 368, 601
\bibitem[\protect\citeauthoryear{Dessart}{2004}]{2004A&A...423..693D} Dessart L., 2004, A\&A, 423, 693
\bibitem[\protect\citeauthoryear{Fullerton et al.}{1997}]{1997A&A...327..699F} Fullerton A.W., 
Massa D.L., Prinja R.K., Owocki S.P., Cranmer S.R., 1997, A\&A, 327, 699
\bibitem[\protect\citeauthoryear{Gayley}{1995}]{1995ApJ...454..410G} Gayley K.G., 1995, ApJ, 454, 410
\bibitem[\protect\citeauthoryear{Hamann}{1981}]{1981A&A....93..353H} Hamann W.-R., 1981, A\&A, 93, 353
\bibitem[\protect\citeauthoryear{Haser et al.}{1995}]{1995A&A...295..136H} Haser S.M., Lennon D.J., Kudritzki R.P., Puls J., Pauldrach A.W.A., Bianchi L.,
Hutchings J.B., 1995, A\&A, 295, 136
\bibitem[\protect\citeauthoryear{Henrichs et al.}{1983}]{1983ApJ...268..807H} Henrichs H.F., Hammerschlag-Hensberge G., Howarth I.D., Barr P., 1983, ApJ, 268, 807
\bibitem[\protect\citeauthoryear{Howarth \& Prinja}{1989}]{1989ApJS...69..527H} Howarth I.D., Prinja R.K., 1989, ApJS, 69, 527
\bibitem[\protect\citeauthoryear{Kaper et al.}{1996}]{1996A&AS..116..257K} Kaper L., Henrichs H.F., Nichols J.S., Snoek L.C., Volten H., Zwarthoed G.A.A.,
1996, A\&AS, 116, 257
\bibitem[\protect\citeauthoryear{Kaper et al.}{1997}]{1997A&A...327..281K} Kaper L., Henrichs H.F., Fullerton A.W., Ando H., Bjorkman K.S., Gies D.R., Hirata R.,
Kambe E., McDavid D., Nichols J.S., 1997, A\&A, 327, 281
\bibitem[\protect\citeauthoryear{Kaper et al.}{1999}]{1999A&A...344..231K} Kaper L., Henrichs H.F., Nichols J.S., Telting J.H., 1999, A\&A, 344, 231
\bibitem[\protect\citeauthoryear{Kee}{2015}]{2015PhDT..........K} Kee N.D., 2015, PhD dissertation, University of Delaware
\bibitem[\protect\citeauthoryear{Krti\v{c}ka \& Kub\'{a}t}{2010}]{2010A&A...519A..50K} Krti\v{c}ka J., Kub\'{a}t J., 2010, A\&A, 519, A50
\bibitem[\protect\citeauthoryear{Lamers, Cerruti-Sola \& Perinotto}{1987}]{1987ApJ...314..726L} Lamers H.J.G.L.M., Cerruti-Sola M., Perinotto M., 1987, ApJ, 314, 726
\bibitem[\protect\citeauthoryear{Lamers \& Leitherer}{1993}]{1993ApJ...412..771L} Lamers H.J.G.L.M., Leitherer C., 1993, ApJ, 412, 771
\bibitem[\protect\citeauthoryear{Marcolino et al.}{2013}]{2013MNRAS.431.2253M} Marcolino W.L.F., Bouret J.-C., Sundqvist J.O., Walborn N.R., Fullerton A.W.,
Howarth I.D., Wade G.A., ud-Doula A., 2013, MNRAS, 431, 2253
\bibitem[\protect\citeauthoryear{Massa \& Prinja}{2015}]{2015ApJ...809...12M} Massa D., Prinja R.K., 2015, ApJ, 809, 12
\bibitem[\protect\citeauthoryear{Mullan}{1986}]{1986A&A...165..157M} Mullan D.J., 1986, A\&A, 165, 157
\bibitem[\protect\citeauthoryear{Owocki et al.}{1988}]{1988ApJ...335..914O} Owocki S.P., Castor J.I., Rybicki G.B., 1988, ApJ, 335, 914
\bibitem[\protect\citeauthoryear{Owocki, Cranmer \& Fullerton}{1995}]{1995ApJ...453L..37O} Owocki S.P., Cranmer S.R., Fullerton A.W., 1995, ApJ, 453, L37
\bibitem[\protect\citeauthoryear{Owocki}{1999}]{1999LNP...523..294O} Owocki S.P., 1999, in \textit{Variable and Non-spherical Stellar Winds
in Luminous Hot Stars}, Proceedings of the IAU Colloquium No. 169 held in Heidelburg, Germany, eds. Wolf B., Stahl O., Fullerton A.W., Lecture Notes
in Physics, 523, 294
\bibitem[\protect\citeauthoryear{Prinja}{1988}]{1988MNRAS.231P..21P} Prinja R.K., 1988, MNRAS, 231, 21
\bibitem[\protect\citeauthoryear{Prinja \& Howarth}{1986}]{1986ApJS...61..357P} Prinja R.K., Howarth I.D., 1986, ApJS, 61, 357
\bibitem[\protect\citeauthoryear{Prinja, Howarth \& Henrichs}{1987}]{1987ApJ...317..389P} Prinja R.K., Howarth I.D., Henrichs H.F., 1987, ApJ, 317, 389
\bibitem[\protect\citeauthoryear{Puls, Owocki \& Fullerton}{1993}]{1993A&A...279..457P} Puls J., Owocki S.P., Fullerton  A.W., 1993, A\&A, 279, 457
\bibitem[\protect\citeauthoryear{Puls, Springmann \& Lennon}{2000}]{2000A&AS..141...23P} Puls J., Springmann U., Lennon M., 2000, A\&AS, 141, 23
\bibitem[\protect\citeauthoryear{Ramiaramanantsoa et al.}{2014}]{2014MNRAS.441..910R} Ramiaramanantsoa T., Moffat A.F.J., 
Chen\'{e} A.-N., Richardson N.D., Henrichs H.F.,
Desforges S., Antoci V., Rowe J.F., Matthews J.M., Kuschnig R., Weiss W.W., Sasselov D., Rucinski S.M., Guenther D.B., 2014, MNRAS, 441, 910
\bibitem[\protect\citeauthoryear{Rauw et al.}{2015}]{2015A&A...580A..59R} Rauw G., Herv\'{e} A., Naz\'{e} Y., Gonz\'{a}lez-P\'{e}rez J.N., Hempelmann A.,
Mittag M., Schmitt J.H.M.M., Schr\"{o}der K.-P., Gosset E., Eenens P., Uuh-Sonda J.M., 2015, A\&A, 580, A59
\bibitem[\protect\citeauthoryear{Repolust, Puls \& Herrero}{2004}]{2004A&A...415..349R} Repolust T., Puls J., Herrero A., 2004, A\&A, 415, 349
\bibitem[\protect\citeauthoryear{Sobolev}{1960}]{1960mes..book.....S} Sobolev V.V., 1960, \textit{Moving Envelopes of Stars} (Cambridge, MA: Harvard Univ. Press)
\bibitem[\protect\citeauthoryear{Sundqvist, Puls \& Feldmeier}{2010}]{2010A&A...510A..11S} Sundqvist J.O., Puls, J., Feldmeier A., 2010, A\&A, 510, A11
\bibitem[\protect\citeauthoryear{Sundqvist et al.}{2012}]{2012MNRAS.423L..21S} Sundqvist J.O., ud-Doula A., Owocki S.P., Townsend R.H.D., Howarth I.D.,
Wade G.A., 2012, MNRAS, 423, L21
\bibitem[\protect\citeauthoryear{Sundqvist, Puls \& Owocki}{2014}]{2014A&A...568A..59S} Sundqvist J.O., Puls J., Owocki S.P., 2014, A\&A, 568, A59
\bibitem[\protect\citeauthoryear{Wade et al.}{2014}]{2014IAUS..302..265W} Wade G.A., Grunhut J., Alecian E., Neiner C., Auri\`{e}re M., Bohlender D.A., David-Uraz A.,
Folsom C., Henrichs H.F., Kochukhov O., Mathis S., Owocki S., Petit V., MiMeS Collaboration, 2014, IAUS, 302, 265

\end{thebibliography}
\end{document}